\documentclass[12pt]{article}
\usepackage[utf8]{inputenc}
\usepackage[english]{babel}
\usepackage{amsmath, amssymb,amsfonts,fullpage}
\usepackage{array}
\usepackage[dvips]{graphicx}
\usepackage{epsfig,float,color,hyperref}
\usepackage{subfigure} % per subfloat
\usepackage{amsthm}
\usepackage[normalem]{ulem}
\usepackage{booktabs}
\usepackage{tabularx}
\usepackage{multirow}

\usepackage{algorithm}
\usepackage{algorithmic}
\usepackage{bm}
\usepackage{amsmath}

\usepackage{verbatim}
\usepackage{bm,latexsym,mathrsfs,setspace, booktabs}
\usepackage{multirow}
\usepackage{pdflscape}
\usepackage[dvipsnames]{xcolor}

\usepackage[title]{appendix}

\usepackage[authoryear]{natbib}

\DeclareMathOperator\arctanh{arctanh}

\usepackage{tikz}
\usepackage[affil-it]{authblk}

\newcommand{\GO}[1]{{\color{red}{GO:} #1}}

\def \ba {\begin{array}}
\def \ea {\end{array}}

\def \Pr {{\mathbb P}}

\newtheorem{remark}{Remark}[section]
\newtheorem{proposition}{Proposition}[section]

\newtheorem{corollary}{Corollary}[section]
\newtheorem{lemma}{Lemma}[section]
\newtheorem{definition}{Definition}[section]
\usepackage{authblk}

\begin{document}

\title{Hybrid Geometry-Adaptive MCMC for Bayesian Inference in Higher-Order Ising Models}
%\title{Hybrid MCMC for Bayesian Inference in the Inverse Ising Model with Three-Body Interactions}

\author[1]{Godwin Osabutey\thanks{gosabutey@unimore.it}}
\author[2]{Robert Richardson\thanks{richardson@stat.byu.edu}}
\author[2]{Garritt L. Page\thanks{page@stat.byu.edu}}
\affil[1]{Department of Physics, Computer and Mathematical Sciences, University of Modena and Reggio Emilia, Modena, Italy}
\affil[2]{Department of Statistics, Brigham Young University, Provo, USA}
\setcounter{Maxaffil}{0}
\renewcommand\Affilfont{\small}

\date{\today}

\maketitle
\begin{abstract}
We address the inverse problem for the mean-field Ising model with two- and three-body interactions using a Bayesian framework. Parameter recovery in this setting is notoriously difficult, particularly near phase transitions, at criticality, and under non-identifiability, where conventional estimators and standard MCMC samplers fail. To overcome these challenges, we develop a hybrid algorithm that combines Adaptive Metropolis Hastings with geometry-aware Riemannian manifold Hamiltonian dynamics. This approach yields substantially improved mixing and convergence in the three-dimensional parameter space. Through simulated experiments across representative regimes, we demonstrate that the method achieves accurate density reconstruction and reliable uncertainty quantification even in settings where existing approaches are unstable or inapplicable. %Beyond statistical mechanics, our results have implications for learning in higher-order energy-based models and Boltzmann machine training.

\end{abstract}

\maketitle


\noindent

\section{Introduction}
The Ising model is a prototypical binary spin system originally developed to describe ferromagnetic phase transitions and has since become a cornerstone for modeling dependent binary data in statistics and network science. Its equilibrium distribution is given by the Gibbs meausre
\begin{equation}\label{gibbs}
    P(\mathbf{x}; \boldsymbol{\theta}) = \frac{\exp\!\big(-\beta H(\mathbf{x};\boldsymbol{\theta})\big)}{Z(\boldsymbol{\theta})},
\end{equation}
where $Z(\boldsymbol{\theta})=\sum_{\mathbf{x}\in\{\pm1\}^N}\exp(-\beta H(\mathbf{x};\boldsymbol{\theta}))$ is the intractable partition function (i.e., normalization constant), $H$ is the Hamiltonian parametrized by $\boldsymbol{\theta}$, and $\beta>0$ is the inverse temperature. We study the cubic (three-body) extension \cite{Contucci_Kertész_Osabutey_2022,Contucci_Osa_Ver_2023,Contucci_Mingione_Osabutey_2023}:
\begin{equation}\label{Cubic_model}
    H(\mathbf{x};\boldsymbol{\theta}) = - \sum_{\langle i,j,k\rangle} K_{ijk}x_i x_j x_k - \sum_{\langle i,j\rangle} J_{ij}x_i x_j - \sum_{i=1}^N h_i x_i,
\end{equation}
so that $\boldsymbol{\theta}=(\{K_{ijk}\},\{J_{ij}\},\{h_i\})$ encodes triadic couplings, pairwise couplings, and local fields.

The inverse problem, recovering these microscopic parameters $\boldsymbol{\theta}$ from macroscopic equilibrium observations such as empirical magnetizations and moments, is central to fields ranging from neuroscience and protein contact prediction to materials science and social systems \citep{Schneidman2006, Mezard_Mora_2009, Torquato_2009, GalloBarra2009}. This task is particularly challenging near criticality, where distributions are strongly non-Gaussian and often multimodal. The general intractability of $Z(\boldsymbol{\theta})$ has spurred a variety of estimation strategies, including pseudolikelihood methods \citep{Chatterjee_2007, Aurell_2012, Bhattacharya_MPLE_2018}, variational Bayes \citep{kim2021variational}, reweighting schemes combined with sequential Monte Carlo \citep{Habeck_Bayes2014}, and analytic mean-field or method-of-moments formulas \citep{Opoku_Osabutey_Kwofie_2019, FedeleVernia2013, Contucci_Osa_Ver_2023}.

A primary motivation for this work lies in its close connection to machine learning via energy-based models (EBMs). Boltzmann machines (originating in the 1980s \citep{Ackley1985, HintonSejnowski1986}) are generative models whose energy functions are directly analogous to physical Hamiltonians. However, recent advances have emphasized the importance of higher-order interactions. For instance, frameworks using restricted Boltzmann machines (RBMs) to accurately recover two- and three-body couplings from data have emerged \citep{DecelleNavasGomezSeoane2025}. Meanwhile, RBMs continue to demonstrate the ability to encode thermodynamic features of the Ising model and identify critical behavior from configuration data \citep{GuZhang2022}. A broader review of EBMs further clarifies their formal and computational connections to statistical mechanics and highlights the need for scalable inference methods in structured generative modeling \citep{Carbone2025}.

In this work, we make three contributions to the Bayesian inference of the three-body Ising model. First, under a mean-field (fully connected) assumption, the partition function admits a low-dimensional representation in terms of the empirical magnetization, which is amenable to Laplace-type asymptotics as $N\to\infty$; we exploit these expansions to derive tractable expressions for the log-likelihood, its gradient, and the Fisher information matrix. Second, recognizing that standard Markov Chain Monte Carlo (MCMC) mixes poorly in the presence of strong parameter correlations and near-non-identifiability, we design a hybrid sampler that alternates between Riemannian-manifold Hamiltonian Monte Carlo (RMHMC) proposals, which use the Fisher metric for geometry-aware moves, and adaptive Metropolis-Hastings (AMH) updates for large, adaptive jumps. This RMAHMC–AMH hybrid combines efficient local exploration with robust global moves. Third, we address practical numerical challenges, including grid-based initialization and regularization of the metric to ensure numerical stability.

We evaluate our methodology on representative regimes where three-body interactions are notoriously difficult to infer: bimodal (phase coexistence), unimodal (well-identified), critical, and near-non-identifiable parameter configurations. Simulation studies demonstrate that the hybrid sampler achieves faster convergence and more reliable uncertainty quantification than standalone adaptive or geometry-aware samplers. Even in non-identifiable regimes, the induced observable distribution is accurately recovered, while individual parameters remain uncertain--a hallmark of sound Bayesian inference.

The rest of this paper is organized as follows: Section~\ref{sec:model-method} develops the model and present the issue of identifiability; Section~\ref{sec:MCMC} how we build the Bayesian model and develops an algorithm to draw samples from the posterior distribution of model parameters; Section~\ref{test} presents challenging scenarios for three-body inference with thorough simulation studies; and Section~\ref{sec:conclusion} concludes. Additional simulation results are provided in Appendix~\ref{AMH RMAHMC}.

\section{The model and methods}\label{sec:model-method}
The network structure considered extends from simple vertex-edge relationships in traditional graph theory to hypergraphs, where triad interactions (faces) play a crucial role. Such higher-order interactions significantly influence system dynamics, often leading to abrupt transitions between different system states \citep{BattistonAmico2021, BensonAbebe2018, SubLeo1999}.
These abrupt transitions, often referred to as the tipping point phenomenon in sociology, represent critical points that trigger substantial changes when crossed \citep{BensonAbebe2018, AlvarezBattiston2021}. Such abrupt phase transitions, exemplified by phenomena like political revolutions \citep{political_rev2015, skocpol1979states}, are of particular interest from a statistical standpoint due to their inherent complexity. Understanding, controlling, and predicting outcomes in networks experiencing abrupt transitions pose unique challenges compared to scenarios involving continuous transitions, such as cultural shifts \citep{hamilton2016cultural, inglehart2020modernization}.  We next detail the mean-field Ising model.

%%% introduce formally the model
\subsection{The mean-field Ising model}
We denote by $x_i \in \{-1,1\}$ the spin variable at vertex $i$, and by $\mathbf{x}=(x_1,\dots,x_N) \in \Omega=\{-1,1\}^N$ a configuration of the network, i.e., the microscopic state of the system. Under the mean-field assumption (i.e., a fully connected interaction graph), all spins are exchangeable and interact symmetrically. The network configuration $\mathbf{x}$ can then be summarized by its empirical mean (i.e., magnetization)
\[
m_N(\mathbf{x}) \;=\; \frac{1}{N}\sum_{i=1}^N x_i \;\in\; [-1,1],
\]
which serves as the macroscopic order parameter. In this setting, isotropic couplings and field are obtained by scaling
\[
K_{ijk} = \frac{K}{3N^2}, \qquad J_{ij} = \frac{J}{2N}, \qquad h_i = h, 
\]
so that the effective parameter vector reduces to $\boldsymbol{\theta}=(K,J,h) \in \mathbb{R}^3$. The normalizations $1/(2N)$ and $1/(3N^2)$ prevent double and triple counting in \eqref{Cubic_model}. Substituting into the Hamiltonian yields the mean-field form
\begin{equation}\label{Cubic mf}
    H(\mathbf{x};\boldsymbol{\theta}) \;=\; -N\left(\frac{K}{3} m_N^3(\mathbf{x}) + \frac{J}{2} m_N^2(\mathbf{x}) + h\, m_N(\mathbf{x}) \right).
\end{equation}
Here, $\boldsymbol{\theta}$ is fixed so that, $K$ describes the interaction strength between all triples of vertex variables, $J$ describes the strength of interaction between all pairs and $h$ is the global field on all vertices. 

The Gibbs distribution \eqref{gibbs} requires evaluating the normalizing constant $Z$, which lacks a closed form. Standard approaches include the Hubbard–Stratonovich (i.e., Gaussian) transform, large-deviation methods \citep{Ellis06,Dembo_Zeitouni_2010}, cluster expansions and others \citep{baxter2016exactly}. For the non-convex Hamiltonian \eqref{Cubic mf}, Gaussian transforms are inefficient; instead, we adopt a microcanonical large-deviation approach (see \cite{baxter2016exactly,Talagrand_2003}):
\begin{corollary}
Let $S_m=\{-1+\tfrac{2k}{N}: k=0,\dots,N\}$ denote the possible values of $m_N(\mathbf{x})$. Then
\begin{equation}\label{normalizaton factor_m}
    Z(\bm\theta) \;=\; \sum_{m\in S_m} A_N(m)\, \exp\!\big(-H(\mathbf{x};\boldsymbol{\theta})\big),
\end{equation}
where $A_N(m)=\binom{N}{\frac{1+m}{2}N} = \text{card}\{\mathbf{x}\in\{-1,1\}^N : m_N(\mathbf{x})=m\}$ counts the microstates corresponding to a given macrostate $m$. Here $\beta$ is absorbed into $\boldsymbol{\theta}$. %Bounds on $A_N(m)$ and asymptotic expansions of $Z(\bm\theta)$ are given in Appendix~\ref{appendix partition}.
\end{corollary}
Thus, instead of summing over $2^N$ configurations, the Gibbs weight reduces to a sum over $N+1$ macrostates. For a configuration $\mathbf{x}$ with empirical mean $m_N(\mathbf{x})=m$, the probability of observing macrostate $m$ is
\begin{equation}\label{likelihood}
    P(m_N(\mathbf{x})=m; \boldsymbol{\theta}) %=  \sum_{\mathbf{x}\in \{-1,1\}^N: m_N(\mathbf{x})=m} \frac{ e^{- H(\mathbf{x};\boldsymbol{\theta})}}{Z(\bm\theta)}
    \;=\; \frac{A_N(m)\,\exp\!\big(-H(\mathbf{x};\boldsymbol{\theta})\big)}{\sum_{m'\in S_m} A_N(m')\,\exp\!\big(-H(\mathbf{x};\boldsymbol{\theta})\big)}.
\end{equation}
This formulation implies that all configuration $\mathbf{x}$ with the same $m$ have equal probability. It also highlights the reduction from a microscopic to a macroscopic likelihood, which is especially relevant for inference in the large-$N$ limit. Define the generating functional,
\[ f_N = \frac{1}{N} \ln Z(\bm\theta) , \]
with thermodynamic limit $f = \lim_{N\to\infty} f_N$, which has been shown to exist \cite{Contucci_Mingione_Osabutey_2023}. Large-deviation analysis yields
\begin{equation}\label{limiting free-enrgy}
    f = \max_{m\in[-1,1]} \Big\{ \tfrac{K}{3} m^3 + \tfrac{J}{2} m^2 + h m - I(m)\Big\},
\end{equation}
where $I(m)$ is the binary entropy rate
\[I(m)=\frac{1-m}{2}\log\left(\frac{1-m}{2}\right)+\frac{1+m}{2}\log\left(\frac{1+m}{2}\right).\] 
The maximizers of $f$ satisfy the mean-field self-consistency relation
\[m \;=\; \tanh(K m^2 + J m + h), \]
which coincides with the expected magnetization (or Gibbs average) $\mathbb{E}[m_N(\mathbf{x})]$ in the unimodal regime.

For an observable $w(\mathbf{x})$, its expectation is
\begin{equation}\label{exp_obs}
    \mathbb{E}[w(\mathbf{x})] \;=\; \sum_{\mathbf{x}\in \Omega} w(\mathbf{x})\, P(m_N(\mathbf{x})=m;\boldsymbol{\theta}).
\end{equation}
In the forward problem, the relevant quantities are low-order moments (e.g., $\mathbb{E}[w(\mathbf{x})]$ and correlations). In the inverse problem of interest here, the goal is to recover the parameters $\boldsymbol{\theta}=(K,J,h)$ given macroscopic observations, assuming full knowledge of $\mathbf{x}$.

%Adopting a Bayesian framework, we place weakly informative Gaussian priors on the parameters:
%\[ K \sim \mathcal{N}(0,2), \quad J \sim \mathcal{N}(0,2), \quad h \sim \mathcal{N}(0,2).\]
%For large $N$, the influence of these priors is negligible compared with the likelihood. Posterior inference is then performed using the advanced MCMC algorithms described in Section~3.
%%%%%%%%%%%%%%%%%%%%

\subsection{Bayesian formulation}
Let $\mathbf{X} = \{ \mathbf{x}^{(1)}, \mathbf{x}^{(2)}, \dots, \mathbf{x}^{(M)} \}$ be a dataset of $M$ independent and identically distributed (i.i.d.) spin configurations, where each $\mathbf{x}^{(i)} \in \{-1, +1\}^N$. Bayesian inference is based on updating prior beliefs about $\bm{\theta}$ in light of the observed data, resulting in the posterior distribution. This update is governed by Bayes' theorem:

\begin{equation}
p(\bm{\theta} | \mathbf{X}) = \frac{p(\mathbf{X} | \bm{\theta}) \, p(\bm{\theta})}{p(\mathbf{X})} \propto p(\mathbf{X} | \bm{\theta}) \, p(\bm{\theta}),
\label{eq:bayes}
\end{equation}
where $p(\bm{\theta} | \mathbf{X})$ is the posterior distribution, $p(\mathbf{X} | \bm{\theta})$ is the likelihood, and $p(\bm{\theta})$ the prior distribution.  
The primary objective of our analysis is to characterize the posterior distribution $p(\bm{\theta} | \mathbf{X})$ employing Markov Chain Monte Carlo (MCMC) methods, to generate samples $\{ \bm{\theta}^{(1)}, \bm{\theta}^{(2)}, \dots, \bm{\theta}^{(S)} \}$ that are asymptotically distributed according to the posterior. These samples are then used for Monte Carlo estimation of posterior summaries, such as means, credible intervals, and marginal densities.

Since we are adopting a Bayesian approach, to finish the model specification based on \eqref{gibbs} with Hamiltonian \eqref{Cubic mf}, we employed the following weakly informative Gaussian priors on $\boldsymbol{\theta}$: $K \sim N(0,2)$, $J \sim N(0,2)$, $h \sim N(0,2)$.  We note however that our experiments are based on a large $N$ making it so that prior specification has little influence on resulting inference.  

%%%%%%%%%%%%%%%%%%%%
\subsection{Identifiability}
This section addresses the issue of non-identifiability inherent in the inverse problem for the Ising model. A model's parameters $\bm{\theta}$ are considered \textit{non-identifiable} if distinct parameter values yield observationally equivalent models. %Formally, let $L(\mathbf{X} ; \bm{\theta})$ be the model likelihood. 
The parameter vector $\bm{\theta}$ is non-identifiable if there exists $\bm{\theta}_1 \neq \bm{\theta}_2$ such that $p(\mathbf{X} | \bm{\theta}_1) = p(\mathbf{X} | \bm{\theta}_2)$ for all possible data $\mathbf{X}$.

In the inverse Ising setting, this typically arises when different combinations of the model parameters $(h,J,K)$ produce identical low-order moments (e.g., mean magnetisation $\mu_1$, susceptibility $\mu_2 - \mu_1^2$). Since the model belongs to an exponential family, the problem of inferring $\bm{\theta}$ from empirical moments is well-posed if and only if the mapping
\begin{equation}\label{identifiability map}
\bm{\theta} \mapsto \bm{\mu}(\bm{\theta})=\mathbb{E}_{\bm{\theta}}\big[s(\mathbf{X})\big],
\qquad 
s(\mathbf{x})=\begin{pmatrix}\tfrac{1}{3}m_N(\mathbf{x})^3\\[2pt]\tfrac{1}{2}m_N(\mathbf{x})^2\\[2pt]m_N(\mathbf{x})\end{pmatrix},
\end{equation}
is injective where $s(\mathbf{x})$ is a vector of the natural sufficient statistics. If this map fails to be injective, then distinct parameter vectors yield identical distributions, and the inverse problem is \emph{structurally non-identifiable}. Equivalently, the preimage $\bm{\mu}^{-1}(\cdot)$ of a given moment vector may contain multiple, distinct points in parameter space.

The other aspect of identifiability issues related to the posedness of the inversion problem happens locally relative to the direction of $\bm\theta$ on the likelihood function.  Let $P_{\bm{\theta}}$ denote the induced distribution on $\Omega$. Suppose $M$ i.i.d.\ observations are drawn from $P_{\bm{\theta}}$, and let $\mathcal{L}(\bm{\theta})$ denote the log-likelihood. Then the Fisher information equals the covariance of the sufficient statistics:
\begin{equation}\label{eqn: fisher information}
I_M(\bm{\theta})=-\mathbb{E}_{\bm{\theta}}\!\left[ \nabla^2_{\bm{\theta}} \mathcal{L}(\bm{\theta})\right]
= M N^2\, \mathrm{Cov}_{\bm{\theta}}\!\big(s(\mathbf{X})\big).
\end{equation}
If $I_M(\bm{\theta})$ is singular (or ill-conditioned), then $\mathcal{L}(\bm{\theta})$ is locally flat in some direction and $\bm{\theta}$ is locally non-identifiable. This happens when small changes made to $\bm\theta$ has no significant effect on the likelihood function. 
\begin{comment}
: small perturbations of $\bm{\theta}$ along that direction produce negligible changes in the likelihood, making stable estimation impossible. 

{\color{red}Writing the moments $\mu_k(\bm{\theta})=\mathbb{E}_{\bm{\theta}}[m_N(\mathbf{X})^k]$, the Fisher information has explicit form:
\[
I_M(\bm{\theta})=M N^2
\begin{pmatrix}
\frac{1}{9}\big(\mu_6-\mu_3^2\big) & \frac{1}{6}\big(\mu_5-\mu_3\mu_2\big) & \frac{1}{3}\big(\mu_4-\mu_3\mu_1\big)\\[6pt]
\frac{1}{6}\big(\mu_5-\mu_3\mu_2\big) & \frac{1}{4}\big(\mu_4-\mu_2^2\big) & \frac{1}{2}\big(\mu_3-\mu_2\mu_1\big)\\[6pt]
\frac{1}{3}\big(\mu_4-\mu_3\mu_1\big) & \frac{1}{2}\big(\mu_3-\mu_2\mu_1\big) & \mu_2-\mu_1^2
\end{pmatrix}.
\]}
\end{comment}
The smallest eigenvalue $\lambda_{\min}(I_M(\bm{\theta}))$ quantifies the weakest constrained direction. Let $\nu$ be a unit vector in the associated eigenspace, then for a small perturbation  $\delta \nu$, $\delta>0$, around $\bm\theta$ the local change in $\mathcal{L}(\bm{\theta})$ is dominated by the quadratic term:
\begin{equation}
\Delta \mathcal{L}(\bm{\theta} + \delta \nu) \; := \; \mathcal{L}(\bm{\theta} + \delta \nu) - \mathcal{L}(\bm{\theta})  \;\approx\; \frac{1}{2}\, \delta^2 \nu^\top \nabla^2_{\bm{\theta}} \mathcal{L}(\bm{\theta})\nu \;\approx\; -\frac{1}{2}\, \delta^2 \nu^\top I_M(\bm{\theta}) \nu,
\end{equation}
so if $\lambda_{\min}(I_M(\bm{\theta})) \ll 1$, many distinct $\bm{\theta}$ produce nearly identical likelihoods. In this model, the principal causes of structural or practical non-identifiability are:
\begin{enumerate}
\item \emph{Symmetries / transformations:} if there exists a nontrivial transformation $T$ on $\Omega$ under which the sufficient statistics $s(\mathbf{x})$ transform in a way that can be absorbed by a reparametrisation of $\bm{\theta}$, then distinct $\bm{\theta}$ may induce the same $P_{\bm{\theta}}$;
\item \emph{Approximate linear dependence of sufficient statistics:} if the components of $s(\mathbf{x})$ are (nearly) linearly dependent under $P_{\bm{\theta}}$ then $\mbox{Cov}_{\bm{\theta}}(s)$ will be nearly singular;
\item \emph{Limited / noisy data:} small $M$ (or large observational noise) reduces the information $I_M(\bm\theta)$ and can render estimation ill-posed in practice.
\end{enumerate}

For the inverse Ising model with two- and three-body terms, identifiability is equivalent to injectivity of $\bm{\theta} \mapsto \bm{\mu}(\bm{\theta})$, or equivalently, positive definiteness of $I_M(\bm{\theta})$. When $I_M(\bm{\theta})$ has very small eigenvalues, likelihood-based inference and standard MCMC sampling become slow or unreliable unless degeneracies are handled explicitly (e.g.\ via reparametrisation, informative priors, or penalisation). We demonstrate this behavior numerically in Section~\ref{test}.
%%%%%%%%%%%%%%%%%%%%%%%%%%%%%%%%%%%%%%%%%

\section{Markov Chain Monte Carlo Algorithm}\label{sec:MCMC}
Sampling from the posterior distribution $p(\bm{\theta} | \mathbf{X})$ for the three-body Ising model is challenging due to strong correlations between parameters. Standard Metropolis-Hastings (MH) sampling exhibits poor mixing. We therefore developed a hybrid sampler combining three advanced MCMC methods: Adaptive Metropolis-Hastings (AMH), Hamiltonian Monte Carlo (HMC), and concepts from the Metropolis-Adjusted Langevin Algorithm (MALA). This hybrid approach falls under the broader class of \textit{coupling techniques} \citep{Jacob_O’Leary_Atchadé_2020, O’Leary_Wang_Jacob_2020}. While all three improved the mixing, the most consistent method found was a combination of all three. 

\subsubsection*{Adaptive Metropolis-Hastings (AMH)}
The AMH algorithm iteratively updates the proposal distribution's covariance matrix using the sampled values at previous iterations. In many scenarios it can intelligently tune the algorithm to force good acceptance rates and therefore good mixing. However, for the Ising model with three-body interaction, the correlation between parameters is physically driven in such a way that acceptance rates were never able to get into acceptable ranges and mixing remained poor, especially when the starting values for the algorithm were not close to the truth. This will be illustrated in Section \ref{test}.

\subsubsection*{Metropolis-Adjusted Langevin Algorithm (MALA)}
MALA leverages the gradient of the log-posterior to make informed proposals, moving the chain toward regions of higher probability. The proposal is given by:
\[
\bm{\theta}^* = \bm{\theta} + \frac{\varepsilon^2}{2} \nabla \log p(\bm{\theta} | \mathbf{X}) + \varepsilon \mathbf{z}, \quad \mathbf{z} \sim \mathcal{N}(0, \mathbf{I}),
\]
where $\varepsilon$ is a step size. To account for parameter correlations, a Riemannian manifold version (RM-MALA) uses a position-dependent metric tensor $\mathbf{G}(\bm{\theta})$ (often the Fisher Information) to precondition the proposals:
\[
\bm{\theta}^* = \bm{\theta} + \frac{\varepsilon^2}{2} \mathbf{G}(\bm{\theta})^{-1} \nabla \log p(\bm{\theta} | \mathbf{X}) + \varepsilon \sqrt{\mathbf{G}(\bm{\theta})^{-1}} \mathbf{z}.
\]
Despite using the geometry of the model parameter space to make more educated proposals, it still resulted in low acceptance rates and bad mixing.  Thus, we don't employ stand alone MALA, but rather use it to improve the performance of hybrid algorithms that we now describe. 

\subsubsection*{Hamiltonian Monte Carlo (HMC)}
HMC introduces auxiliary momentum variables $\mathbf{p} \sim \mathcal{N}(\mathbf{0}, \mathbf{M})$, mass matrix \(\mathbf{M}\succ0\), and uses Hamiltonian dynamics to propose distant states with high acceptance. The dynamics are simulated using the leapfrog integrator with step size $\varepsilon$ and $L$ steps:
\begin{align*}
    \mathbf{p}(\tau + \varepsilon/2) &= \mathbf{p}(\tau) + \frac{\varepsilon}{2} \nabla_{\bm{\theta}} \log p(\bm{\theta}(\tau) | \mathbf{X}), \\
    \bm{\theta}(\tau + \varepsilon) &= \bm{\theta}(\tau) + \varepsilon \mathbf{M}^{-1} \mathbf{p}(\tau + \varepsilon/2), \\
    \mathbf{p}(\tau + \varepsilon) &= \mathbf{p}(\tau + \varepsilon/2) + \frac{\varepsilon}{2} \nabla_{\bm{\theta}} \log p(\bm{\theta}(\tau + \varepsilon) | \mathbf{X}),
\end{align*}
where $\tau$ is a placeholder describing the proposal path. The mass matrix $\mathbf{M}$ is crucial for performance. This approach does improve the acceptance rates for sampling posterior draws for the Ising model parameters. However, it often got stuck in flat areas of the likelihood. We use the Riemannian Manifold Adjusted HMC (RMAHMC) extension \citep{Girolami_Calderhead_2011}, which sets $\mathbf{M} = \mathbf{G}(\bm{\theta})$, allowing the algorithm to adapt to the local geometry of the target distribution. This adaptation allows for efficient sampling of high-dimensional parameter spaces, making it particularly useful in our study. 

This combination of MALA and HMC sampling was much more consistent than either on its own and led to good sampling, but it often took several millions of draws to converge. To speed up the algorithm, we adopted a hybrid model with adaptive Metropolis-Hastings. Essentially, the algorithm alternates between a RMAHMC draw and a AMH draw. The RMAHMC draws have a more controllable acceptance depending on $\varepsilon$ but moves very slowly and sometimes even in the wrong direction in certain parts of the likelihood while searching for the correct posterior space  for convergence. The AMH draws can make larger jumps which are almost always rejected, but in many cases have helped find posterior convergence in a fraction of the samples. 

\subsubsection*{Hybrid RMAHMC-AMH Sampler}
Neither AMH, MALA, nor HMC alone provided sufficiently efficient and reliable sampling. We therefore designed a hybrid sampler that alternates between RMAHMC and AMH steps. The rationale is twofold:
\begin{enumerate}
    \item The RMAHMC steps use geometric information to efficiently explore the posterior locally but can be slow to traverse between distant modes.
    \item The AMH steps, while often rejected, can occasionally make large jumps that significantly advance exploration, especially in the early stages of sampling or in flat likelihood regions.
\end{enumerate}
This combination leverages the strengths of both methods, leading to more consistent convergence. The complete procedure is detailed in Algorithm \ref{algorithm}.

\begin{algorithm}[h]
\caption{Hybrid RMAHMC-AMH Sampler}\label{algorithm}
\begin{algorithmic}[1]
\STATE \textbf{Input:} Initial state $\bm{\theta}_0$, total samples $N$, leapfrog step size $\varepsilon$, number of leapfrog steps $L$, adaptation scaling factor $\gamma$.
\STATE \textbf{Initialize:} Set $\bm{\theta} \leftarrow \bm{\theta}_0$, empirical covariance $\mathbf{C} \leftarrow \mathbf{I}$.
\FOR{$i=1$ to $N$}
    \IF{$i$ is odd}
        \STATE \textbf{Riemannian Manifold HMC Step}
        \STATE Compute gradient $\mathbf{g} \gets \nabla \log p(\bm{\theta} | \mathbf{X})$
        \STATE Compute metric tensor $\mathbf{G} \gets -\mathbb{E}[\nabla^2 \log p(\bm{\theta} | \mathbf{X})]$ \COMMENT{See Eq. (\ref{eqn: fisher information})}
        \STATE Sample momentum $\mathbf{p} \sim \mathcal{N}(\mathbf{0}, \mathbf{G})$
        \STATE Set $\bm{\theta}^{(0)} \gets \bm{\theta}$, $\mathbf{p}^{(0)} \gets \mathbf{p}$
        \FOR{$l=0$ to $L-1$ (Leapfrog Integration)} 
            \STATE $\mathbf{p}^{(l + 1/2)} \gets \mathbf{p}^{(l)} + \frac{\varepsilon}{2} \nabla \log p(\bm{\theta}^{(l)} | \mathbf{X})$
            \STATE $\bm{\theta}^{(l+1)} \gets \bm{\theta}^{(l)} + \varepsilon \, \mathbf{G}^{-1} \mathbf{p}^{(l + 1/2)}$
            \STATE $\mathbf{p}^{(l+1)} \gets \mathbf{p}^{(l + 1/2)} + \frac{\varepsilon}{2} \nabla \log p(\bm{\theta}^{(l+1)} | \mathbf{X})$
        \ENDFOR
        \STATE Propose $\bm{\theta}^{*} \gets \bm{\theta}^{(L)}$
        \STATE Compute acceptance probability $\alpha$ using Hamiltonian dynamics
        \STATE Accept or reject $\bm{\theta}^{*}$ with probability $\min(1, \alpha)$
    \ELSE
        \STATE \textbf{Adaptive Metropolis-Hastings Step}
        \STATE Update empirical covariance $\mathbf{C}$ using history of chain
        \STATE Propose $\bm{\theta}^{*} \sim \mathcal{N}(\bm{\theta}, \gamma \mathbf{C})$
        \STATE Compute acceptance probability $\alpha = \frac{p(\bm{\theta}^{*} | \mathbf{X})}{p(\bm{\theta} | \mathbf{X})}$
        \STATE Accept or reject $\bm{\theta}^{*}$ with probability $\min(1, \alpha)$
    \ENDIF
    \STATE Set $\bm{\theta}_i \gets \bm{\theta}$ \COMMENT{Update current state}
\ENDFOR
\end{algorithmic}
\end{algorithm}
%%%%%%%%%%%%%%%%%%%%%%%%

\subsubsection*{Gradient and metric tensor calculation}\label{sec: metric tensor}
%Let $\bm{\mu}(\bm{\theta}) = (\mu_1, \mu_2, \mu_3)^T$ be the vector of theoretical moments $\mathbb{E}[m_N(\mathbf{x})], \mathbb{E}[m_N(\mathbf{x})^2], \mathbb{E}[m_N(\mathbf{x})^3]$ of the Gibbs distribution (Eq. \ref{gibbs}). For a dataset of $M$ independent configurations $\{\mathbf{x}^{(i)}\}_{i=1}^M$, the empirical moments are $\bm{\hat{\mu}} = \frac{1}{M}\sum_{i=1}^M (m_i, m_i^2, m_i^3)^T$, where $m_i = m_N(\mathbf{x}^{(i)})$.

Let $m_i:=m_N(\mathbf{x}^{(i)})$, and define empirical and theoretical moments respectively as
\begin{equation}
\hat\mu_k=\frac{1}{M}\sum_{i=1}^M m_i^k,\qquad \mbox{and} \qquad
\mu_k(\bm{\theta})=\mathbb{E}_{\bm{\theta}}\big[m_N(\mathbf{X})^k\big].
\end{equation}
With parameter ordering $\bm{\theta}=(K,J,h)^\top$ and the natural sufficient statistic defined in \eqref{identifiability map} the log-likelihood has the form
\[
\mathcal{L}(\bm{\theta})=N\sum_{i=1}^M \bm{\theta}^\top s(\mathbf{x}^{(i)}) - M\log Z_N(\bm{\theta}).
\]
Hence the score (gradient) satisfies
\[
\nabla_{\bm\theta} \mathcal{L}(\bm{\theta})
= N\sum_{i=1}^M s(\mathbf{x}^{(i)}) - M N\,\mathbb{E}_{\bm{\theta}}[s(\mathbf{X})]
= N M\big(\widehat s - \mathbb{E}_{\bm{\theta}}[s(\mathbf{X})]\big),
\]
with its components expressed via raw moments as,
\[
\nabla_{\bm\theta} \mathcal{L}(\bm\theta)
= N M
\begin{pmatrix}
\dfrac{1}{3}\big(\hat\mu_3-\mu_3(\bm\theta)\big)\\[8pt]
\dfrac{1}{2}\big(\hat\mu_2-\mu_2(\bm\theta)\big)\\[8pt]
\hat\mu_1-\mu_1(\bm\theta)
\end{pmatrix}.
\]
Dividing by $M$ yields the per-samplers score $N(\widehat s-\mathbb{E}_{\bm\theta}[s(\mathbf{X})])$.

Setting $\mu_k=\mu_k(\bm\theta)$, the metric tensor (expected information for the full sampler), which is the scaled covariance of the sufficient statistics, has the explicit entries:
\begin{equation}\label{fisher entries}
G(\bm\theta) =M N^2 %I_M(\bm\theta)=M N^2
\begin{pmatrix}
\frac{1}{9}\big(\mu_6-\mu_3^2\big) & \frac{1}{6}\big(\mu_5-\mu_3\mu_2\big) & \frac{1}{3}\big(\mu_4-\mu_3\mu_1\big)\\[6pt]
\frac{1}{6}\big(\mu_5-\mu_3\mu_2\big) & \frac{1}{4}\big(\mu_4-\mu_2^2\big) & \frac{1}{2}\big(\mu_3-\mu_2\mu_1\big)\\[6pt]
\frac{1}{3}\big(\mu_4-\mu_3\mu_1\big) & \frac{1}{2}\big(\mu_3-\mu_2\mu_1\big) & \mu_2-\mu_1^2
\end{pmatrix}.
\end{equation}

An empirical estimator is
\begin{equation}\label{fisher estimator}
\widehat G(\bm\theta)=M N^2\Big(\frac{1}{M}\sum_{i=1}^M s(\mathbf{x}^{(i)})s(\mathbf{x}^{(i)})^\top - \widehat s\,\widehat s^\top\Big)+\gamma \mathbf{I},
\end{equation}
with \(\gamma>0\) a small regulariser.

Let $G(\bm\theta)\succcurlyeq 0$, then we say $G(\bm\theta)$ is \emph{numerically singular} at tolerance $\delta>0$ if $\lambda_{\min}\big(G(\bm\theta)\big)<\delta$. In this regime proposals using $G(\bm\theta)^{-1}$ become unreliable.
{\it From equation \eqref{fisher estimator} above the following points hold:
\begin{enumerate}
  \item The matrix $\mathbf{I}$ denotes the identity matrix in $\mathbb{R}^{3\times 3}$ (since $\dim(\bm\theta)=3$). The additive term $\gamma \mathbf{I}$ (with $\gamma > 0$ small) acts as a ridge regulariser, ensuring positive definiteness and numerical stability when inverting $\widehat G(\bm\theta)$ in algorithms such as Riemannian MALA or HMC.

  \item The inner term
  \[
  \frac{1}{M}\sum_{i=1}^M s(\mathbf{x}^{(i)})s(\mathbf{x}^{(i)})^\top - \widehat s\,\widehat s^\top
  \]
  is precisely the empirical covariance of the sufficient statistics $s(\mathbf{x})$. Therefore, $\widehat G(\bm\theta)$ is nothing but the empirical Fisher information matrix (scaled by $MN^2$), with the $\gamma \mathbf{I}$ correction ensuring invertibility.
\end{enumerate}
}

The scaling factor, regulariser, used was $\gamma = 1/3$ for the AMH proposal covariance and determined empirically for this model to achieve reasonable acceptance rates and may require tuning for other applications.
%%%%%%%%%%%%%%%%%%%%%%%%%%%%%%%%%%%%%%%%%

\subsection{Computational considerations}\label{computational consideration}
Despite the improved performance that the hybrid RMAHMC-AMH sampler exhibit, there are still a few of computational considerations that need to be addressed. For example,  the geometry of the likelihood can make it so that the starting point for the Monte Carlo algorithm can result in poor behavior if not chosen intelligently.  To aid in this process, we propose a simple grid search on the likelihood to determine a starting value. First, we define a coarse grid on the space of possible values for  $K$, $J$, and $h$, evaluating the  likelihood at each combination of these parameters. The algorithm begins at the parameter values associated with the largest likelihood. This procedure improves the probability of converging quickly.

\begin{definition}[Coarse-grid initialization]
Let $\bm{\Theta}=\Theta_K\times\Theta_J\times\Theta_h\subset\mathbb{R}^3$ be a bounded box for $\bm\theta=(K,J,h)^\top$. Fix grid resolutions $\Delta_K,\Delta_J,\Delta_h>0$ and define finite grids
\begin{multline}
\mathcal{G}_K=\{K_{\min}+a\Delta_K: a=0,\dots,A\},\quad
\mathcal{G}_J=\{J_{\min}+b\Delta_J: b=0,\dots,B\},\\
\mathcal{G}_h=\{h_{\min}+c\Delta_h: c=0,\dots,C\}.
\end{multline}
Set $\mathcal{G}=\mathcal{G}_K\times\mathcal{G}_J\times\mathcal{G}_h\subset\bm\Theta$. The grid initializer is
\[
\bm\theta^{(0)} \in \arg\max_{\bm\theta\in\mathcal{G}} \; \mathcal{L}(\bm{\theta}).
\]
\end{definition}

This finite optimization has cost $O(|\mathcal{G}|)$ evaluations of $\mathcal{L}$ and provides a data-driven starting point in a high-density region, empirically reducing burn-in. In large-scale simulation studies one may reuse $\mathcal{G}$ across datasets to amortize cost.

To stabilise geometry during exploration, we use a positive definite surrogate $G_{\chi,\gamma}(\bm\theta)$ defined by
\begin{equation}\label{eq:shrink-metric}
G_{\chi,\gamma}(\bm\theta)\;=\;\underbrace{\Big(\chi\,G(\bm\theta)+(1-\chi)\,\mathrm{diag}\,G(\bm\theta)\Big)}_{\text{off-diagonal shrinkage}} \;+\; \gamma \mathbf{I},
\qquad \chi\in[0,1],\ \gamma>0,
\end{equation}
where $\mathrm{diag}\,G(\bm\theta)$ take only the diagonal entries of $G(\bm\theta)$ and setting all off-diagonal entries to zero. To numerically address this, we reduce the off-diagonals by $(1-\chi)100\%$ where $\chi$ is some small value, such as 0.0001. Surprisingly, even this small adjustment can affect mixing in a negative way. The issue of computational singularity does not affect the model when it has converged to the posterior, so to help with mixing and avoid computational errors, we start the algorithm with $\chi = 0.0001$ and then reduce to 0 after some burn-in. 

\begin{lemma}[SPD preservation]
If $G(\bm\theta)\succ0$ and $\gamma\ge 0$, then $G_{\chi,\gamma}(\bm\theta)\succ0$ for all $\chi\in[0,1]$. Moreover, $G_{1,0}(\bm\theta)=G(\bm\theta)$ and $G_{0,\gamma}(\theta)=\mathrm{diag}\,G(\bm\theta)+\gamma \mathbf{I}$.
\end{lemma}
\begin{proof}[Proof.]
$G_{\chi,0}$ is a convex combination of two semi-positive definite (SPD) matrices $G(\bm\theta)$ and $\mathrm{diag}\,G(\bm\theta)$, hence SPD. Adding $\gamma \mathbf{I}$ preserves SPD.
\end{proof}

In RMAHMC, we replace $G(\bm\theta)$ by $G_{\chi,\gamma}(\bm\theta)$ in both the covariance (or mass) and drift terms. This guarantees well-defined Gaussian draws and numerically stable linear solves even when the unregularised $G(\bm\theta)$ is near-singular (typical away from high-density regions or under moment-estimation noise). The Bayesian MCMC approach here avoids (i) explicit inversion of macroscopic relations to express $(K,J,h)$ in terms of observables \citep{Contucci_Osa_Ver_2023}, and (ii) access to individual microstates in variational estimators \citep{kim2021variational}, since one can target the full posterior $p(\bm\theta|\mathbf{x})$ and compute $G(\bm\theta)$ either analytically or via Monte Carlo with regularisation \eqref{eq:shrink-metric}.

\section{Simulations Based on Challenging \texorpdfstring{$(K,J,h)$}{(K,J,h)}}\label{test}
We now illustrate the performance of the algorithms introduced in Section \ref{sec:MCMC} under estimation scenarios that are known to be challenging in the literature \citep{Contucci_Osa_Ver_2023, FedeleVernia2013, DecelleR2016, Fedele_Vernia_2017}. Our experiments not only enable a direct comparison to previous results, but also extend the analysis to parameter values near the critical point, where issues of non-identifiability become prominent.  

The recovery problem for parameters $\bm{\theta}=(K,J,h)$ is strongly influenced by the structure of the limiting free energy \eqref{limiting free-enrgy}. When this function admits multiple maximizers, the likelihood surface becomes bimodal, reflecting phase coexistence. A unique maximizer corresponds to well-identified parameters, while nearly singular Fisher information produces flat directions and weak identifiability.  

Across all experiments, we run the AMH, RMAHMC, and hybrid RMAHMC--AMH samplers for $5{,}000$ iterations, initializing values using the grid-search procedure described in Section \ref{computational consideration} with increments of $0.2$ per coordinate. In two scenarios, $(K,J,h)=(0,1.2,0)$ and $(K,J,h)=(0,1,0)$, the true values coincide with the search grid, yet chains are deliberately initialized away from the truth to assess robustness. We consistently observe that AMH often becomes trapped in flat regions, leading to slow mixing even after convergence; RMAHMC converges reliably but requires hundreds of thousands of iterations to mix efficiently; and the hybrid RMAHMC--AMH method converges more rapidly and maintains efficient mixing thereafter.  

Data are generated with $N=300$ vertices and $M=1000$ independent configurations $\{\mathbf{x}^{(i)}\}_{i=1}^M$ sampled from the Gibbs distribution with parameter $\bm{\theta}$. While larger systems yield qualitatively similar outcomes, the finite-size effects at $N=300$ highlight differences in sampler behavior more clearly. Although $5{,}000$ iterations do not suffice for full convergence in practice, they reveal distinct convergence patterns. For applied inference, substantially longer chains are recommended.  

\subsection{Case 1: Bimodal Likelihood Surface} \label{sec:bimodal}
Multimodality in the distribution of magnetization $m_N(\mathbf{x})$ arises when the likelihood function \eqref{likelihood} admits multiple local maxima, corresponding physically to coexistence of equilibrium states and mathematically to multiple maximizers of the free energy \eqref{limiting free-enrgy} \citep{Nguyen_Zecchina_Berg_2017, Fedele_Vernia_2017}. Such scenarios have been tackled using clustering-based estimation \citep{DecelleR2016, Rodriguez_Laio_2014, Nguyen_Berg_2012}, which partitions samples into basins of attraction before local fitting, or by exploiting spin-flip symmetry $P(\mathbf{x};\bm{\theta})=P(-\mathbf{x};\bm{\theta})$ \citep{Ito_Kohring_1994}. Both strategies, however, are restrictive: clustering can be computationally intensive, while spin-flip symmetry applies only to special parameter choices.  

In contrast, the Bayesian MCMC framework of Section \ref{sec:MCMC} recovers true parameters without such preprocessing. We illustrate this with two representative scenarios: $(K,J,h) = (0,1.2,0)$, corresponding to the two-body Ising model in the low-temperature regime previously addressed with spin-flip methods \citep{Fedele_Vernia_2017}, and $(K,J,h) = (1.67,0.01,0.1)$, previously estimated via density clustering across $50$ independent datasets \citep{Contucci_Osa_Ver_2023}. Our Bayesian method achieves recovery from a single dataset ($M=1000$ samples), demonstrating robustness with less replication.  

\begin{figure}[H]
    \centering
    \makebox[\textwidth][c]{ % Centering and allowing overflow
        \subfigure[(0,1.2,0)]{\label{LLJ1.2}
            \includegraphics[width=8.9cm]{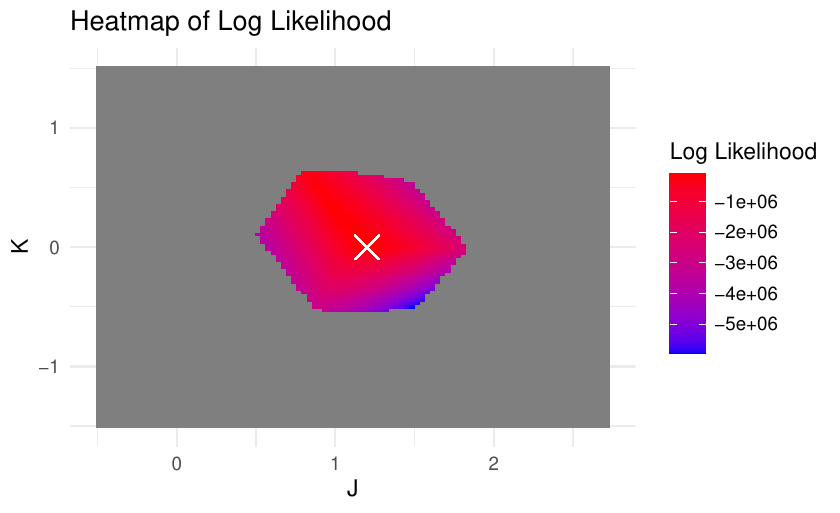}
        }
        \subfigure[(1.67,0.01,0.1)]{\label{LLK1.67}
            \includegraphics[width=8.9cm]{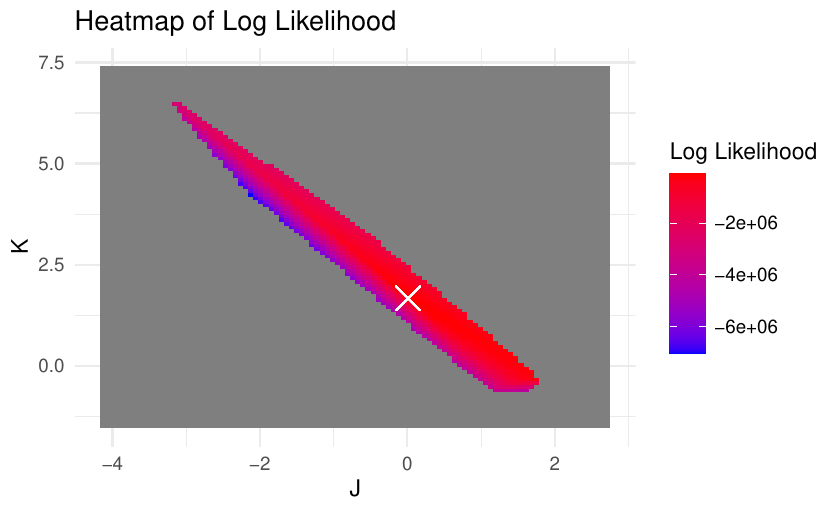}
        }
    }
    \caption{Plot (a) and (b) displays the log-likelihood as a function of $K$ and $J$ while keeping $h$ fixed. The white `x' marker indicates the specific points (i.e., $(0, 1.2, 0)$ and $ (1.67,0.01,0.1)$) in the parameter space we aim to recover. The gray phase corresponds to areas in the parameter space where the likelihood is numerically zero.} 
\end{figure}

\begin{figure}[H]
    \centering
    \makebox[\textwidth][c]{ % Allow images to extend beyond default width
        \begin{minipage}{1.25\textwidth}
            \centering
            % First row (3 images)
            \begin{minipage}{0.32\textwidth}
                \centering
                \subfigure[AMH]{\label{AMH_J1.2}
                    \includegraphics[width=\textwidth]{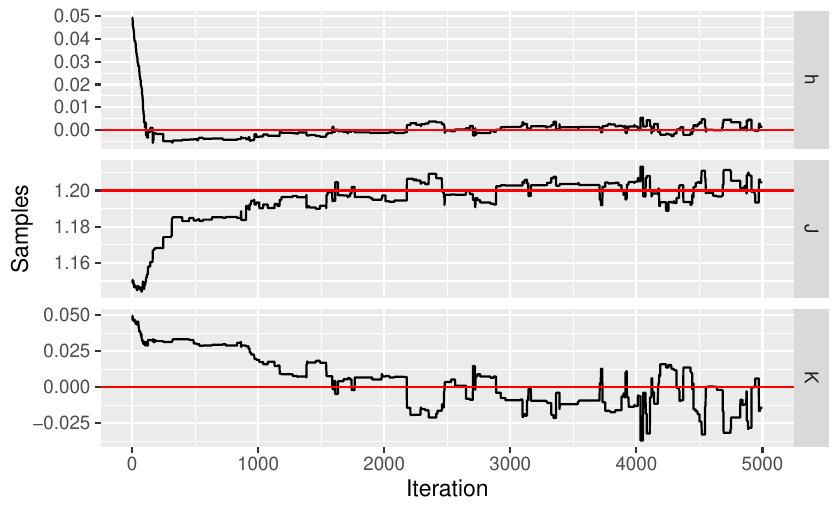}
                }
            \end{minipage}
            \begin{minipage}{0.32\textwidth}
                \centering
                \subfigure[RMAHMC]{\label{RMAHMC_J1.2}
                    \includegraphics[width=\textwidth]{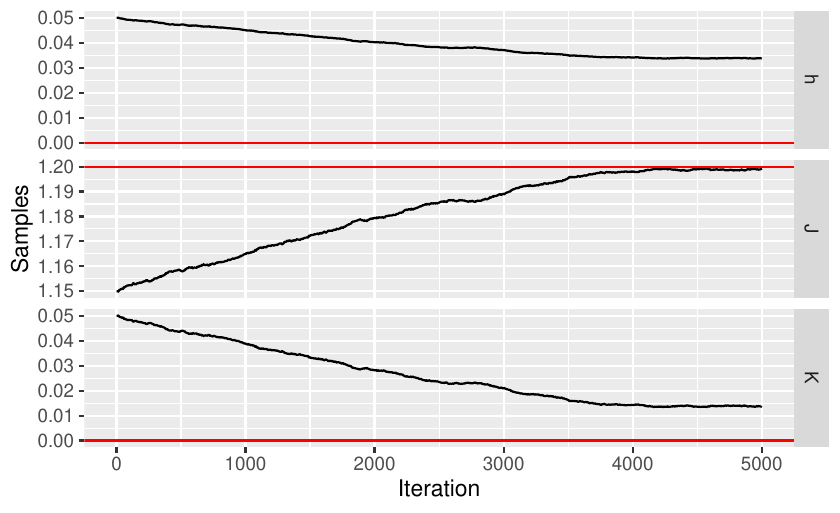}
                }
            \end{minipage}
            \begin{minipage}{0.32\textwidth}
                \centering
                \subfigure[RMAHMC-AMH]{\label{hybrid-K1.2}
                    \includegraphics[width=\textwidth]{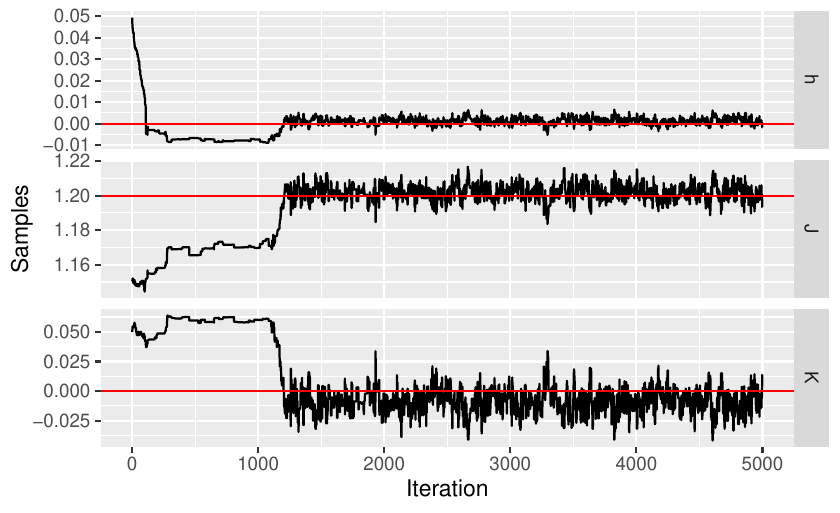}
                }
            \end{minipage} \\[1em] % Space between rows

            % Second row (3 images)
            \begin{minipage}{0.32\textwidth}
                \centering
                \subfigure[AMH]{\label{AMH_K1.67}
                    \includegraphics[width=\textwidth]{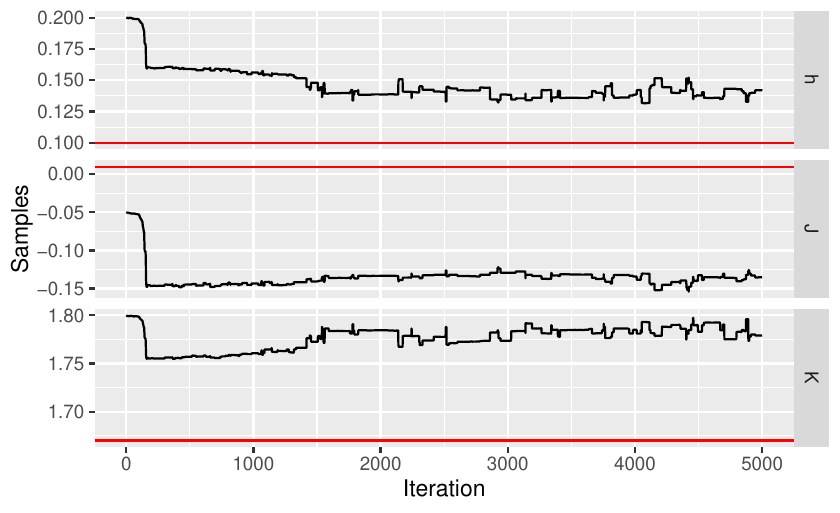}
                }
            \end{minipage}
            \begin{minipage}{0.32\textwidth}
                \centering
                \subfigure[RMAHMC]{\label{RMAHMC_K1.67}
                    \includegraphics[width=\textwidth]{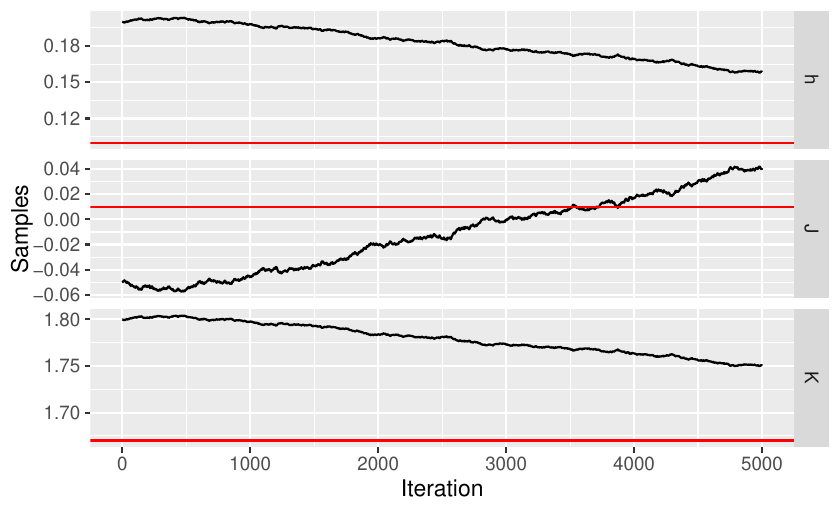}
                }
            \end{minipage}
            \begin{minipage}{0.32\textwidth}
                \centering
                \subfigure[RMAHMC-AMH]{\label{RMAHMC-AMH_K1.67}
                    \includegraphics[width=\textwidth]{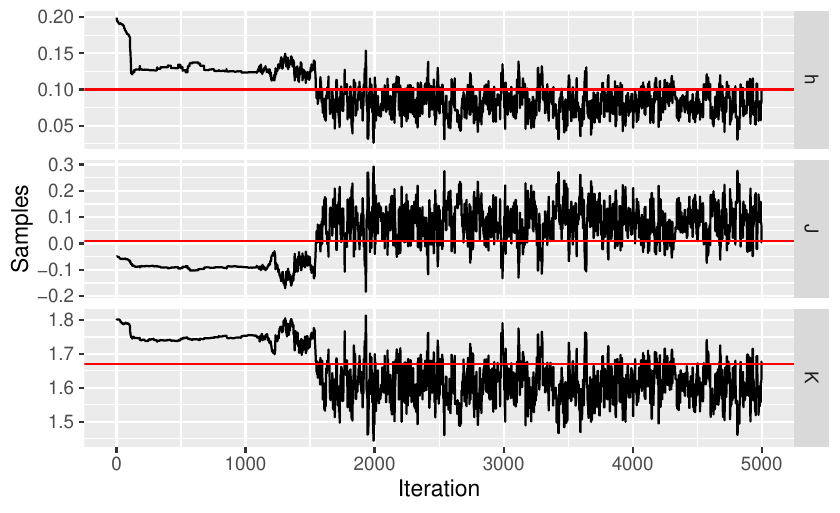}
                }
            \end{minipage}
        \end{minipage}
    }
    \caption{Trace plot of the MCMC samples collected from the posterior distribution of $(K,J,h)$ based on $N = 300$ and $M=1000$ sampled configurations using $(K,J,h)=(0,1.2,0)$ (first row) and $(K,J,h)=(1.67,0.01,0.1)$ (second row).}
    \label{TraceJ=1.2K=1.67}
\end{figure}

Figures \ref{LLJ1.2} and \ref{LLK1.67} show the log-likelihood surfaces for these scenarios. The irregular geometry of the likelihood functions highlights the challenge of constructing efficient samplers. Figure \ref{TraceJ=1.2K=1.67} displays representative MCMC traces for $(K,J,h)$ across the AMH, RMAHMC, and hybrid RMAHMC--AMH algorithms, illustrating the relative efficiency of the hybrid scheme.
%%%%%%%%%%%%

%%%%%%%%%%%%%%%%%%%%%%%%%%%%%
Focusing first on the left column of Figure \ref{TraceJ=1.2K=1.67} (the AMH algorithm), notice the poor mixing in both cases and the very slow convergence. For the middle column (i.e., the RMAHMC algorithm),  convergence is also very slow requiring hundreds of thousands of MCMC iterations before convergence is met. The hybrid RMAHMC-AMH on the other hand converges very quickly (less than 2,000 iterations) and displays very good mixing.  We discuss convergence diagnostics based on the Gelman-Rubin statistic in more detail for all scenarios and algorithms in Section \ref{convergence.measures}.

\begin{figure}[H]
\centering
    \makebox[\textwidth][c]{ % Centering and allowing overflow
        \subfigure[$(0,1.2,0)$]{ \label{GibbsJ1.2}
            \includegraphics[width=7.9cm]{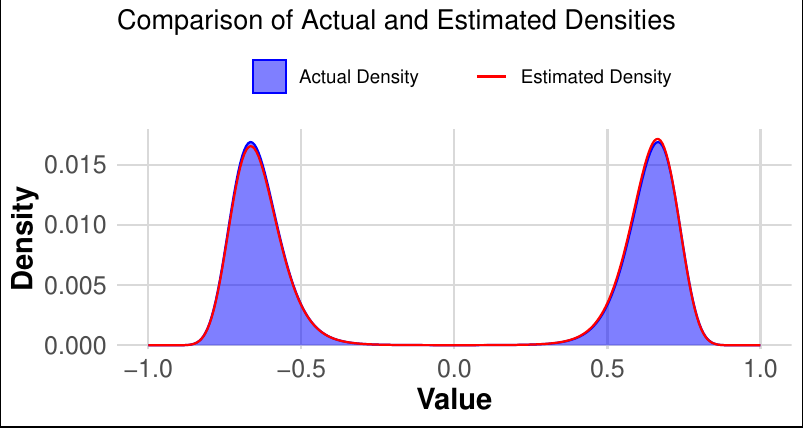}
        }
        \subfigure[$(1.67,0.01,0.1)$]{\label{GibbsK1.67}
            \includegraphics[width=7.9cm]{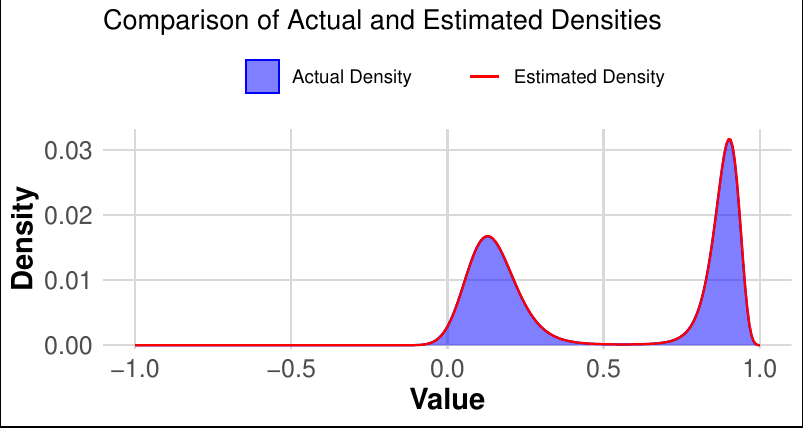}
        }
    }
\caption{Boltzmann-Gibbs distribution of $m_N(\mathbf{x})$ at $N = 300$ with $M = 1000$ for the true and estimated parameters at (a) $(K,J,h) =(0,1.2,0)$ and (b) $(K,J,h)=(1.67,0.01,0.1)$. In (a), the peaks of the distribution are centered around two distinct and opposite values of $m_N(\mathbf{x})$ such that $\mathbb{E}m_N(\mathbf{x})=0$. The two peaks corresponds to global stable states of the system. The larger peak in (b) corresponds to a metastable state with probability approaching $0$ as $N$ increases. The red curve corresponds to the density plot estimated using MCMC samples collected based on the RMAHMC-AMH algorithm.} \label{density bi}
\end{figure}

We next  construct the density of $m_N(\mathbf{x})$ in both scenarios using values for $\boldsymbol{\theta} = (K,J,h)$ estimated based on the MCMC samples collected.  Figure \ref{GibbsJ1.2} corresponds to the $(K,J,h) = (0,1.2,0)$ case, illustrating the two global stable states, where the red curve is obtained using the posterior mean of $\boldsymbol{\theta}$.  For the $(K,J,h) = (1.67,0.01,0.1)$ case, notice that the equilibrium behavior of the system emerges to a metastable state with two peaks for smaller values of $N$ (see Figure \ref{GibbsK1.67}). 

\subsection{Case 2: Unimodal Density} \label{sec:unimodal}
We now consider scenarios in which the distribution of the magnetization $m_N(\mathbf{x})$ is unimodal. Two representative parameter settings are examined: a regular case $\bm{\theta}=(0.5,0.3,0.1)$ and the critical case $\bm{\theta}=(0,1,0)$, the latter marking the onset of a phase transition \citep{Contucci_Osa_Ver_2023}. Recovery of the first configuration is relatively straightforward and serves as a benchmark, while the second is considerably more delicate, since at $(K,J,h)=(0,1,0)$ the susceptibility and third moment of $m_N(\mathbf{x})$ diverge, complicating both the likelihood surface and its numerical exploration.

\begin{figure}[H]
\centering
    \makebox[\textwidth][c]{ % Centering and allowing overflow
        \subfigure[$(0.5,0.3,0.1)$]{\label{LLU 1}
            \includegraphics[width=7.9cm]{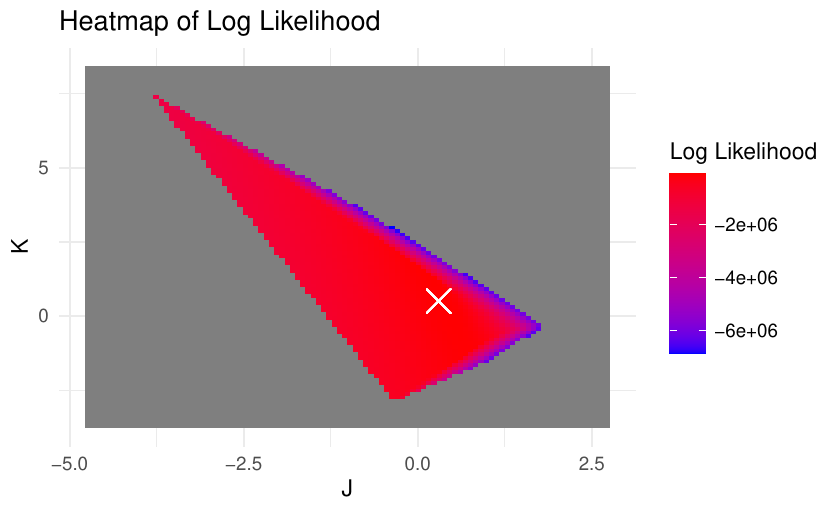}
        }
        \subfigure[$(0,1,0)$]{\label{LLU crit}
            \includegraphics[width=7.9cm]{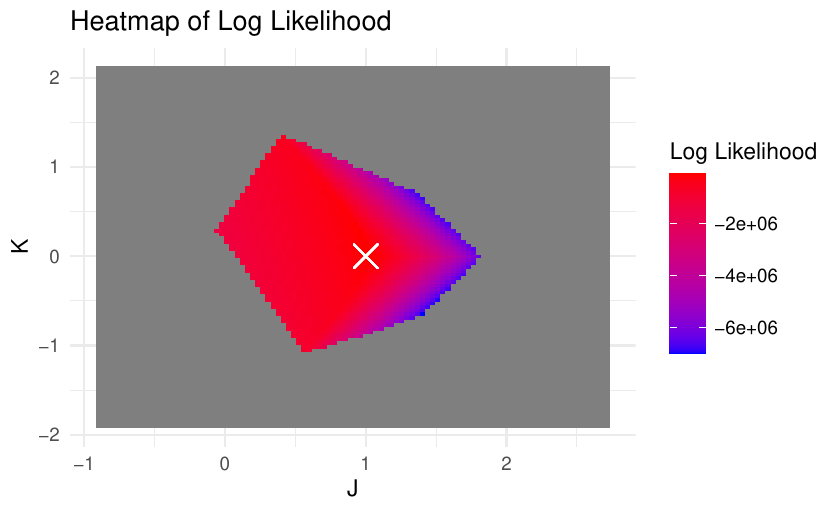}
        }
    }
\caption{Plot (a) displays the log-likelihood as a function of $K$ and $J$ for fixed $h=0.1$. Plot (b) displays the log-likelihood as a function of $K$ and $J$ while keeping $h$ fixed at 0. The white `x' marker indicates the point in the parameter space we aim to recover. The gray phase corresponds to areas in the parameter space where the likelihood is numerically zero and the red area depicts the flat regions of the likelihood function.}
\end{figure}

Figures \ref{LLU 1} and \ref{LLU crit} present the log-likelihood landscapes for the two cases. As in the bimodal setting (Section \ref{sec:bimodal}), irregular geometries persist, posing difficulties for posterior sampling.  
%%%%%%%%%%%%%%%%%%%%%%%%%%%%%
%%%%%%%%%%%%%%%%%%%%%%%%%%%

\begin{figure}[H]
    \centering
    \makebox[\textwidth][c]{ % Allow images to extend beyond default width
        \begin{minipage}{1.2\textwidth}
            \centering
            % First row (3 images)
            \begin{minipage}{0.32\textwidth}
                \centering
                \subfigure[]{\label{uniAHMC}
                    \includegraphics[width=\textwidth]{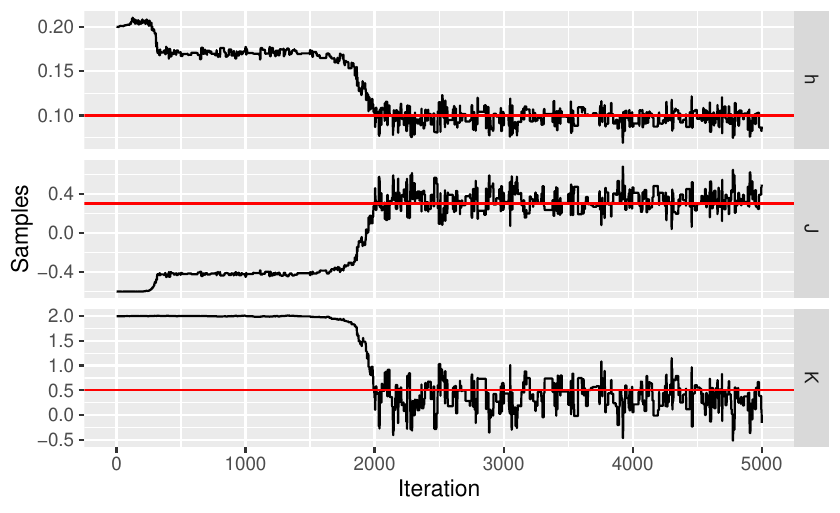}
                }
            \end{minipage}
            \begin{minipage}{0.32\textwidth}
                \centering
                \subfigure[]{\label{uniRMAHMC}
                    \includegraphics[width=\textwidth]{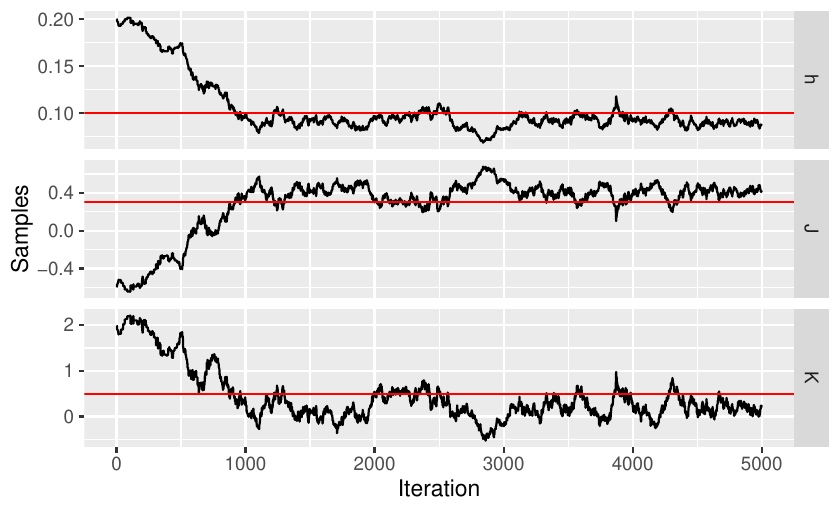}
                }
            \end{minipage}
            \begin{minipage}{0.32\textwidth}
                \centering
                \subfigure[]{\label{uniHybrid}
                    \includegraphics[width=\textwidth]{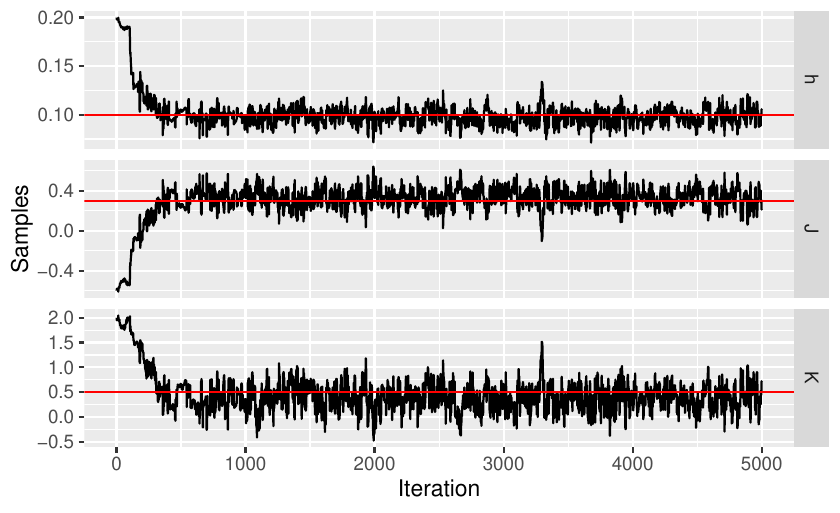}
                }
            \end{minipage} \\[.5em] % Space between rows

            % Second row (3 images)
            \begin{minipage}{0.32\textwidth}
                \centering
                \subfigure[AMH]{\label{uniAHMC_crit}
                    \includegraphics[width=\textwidth]{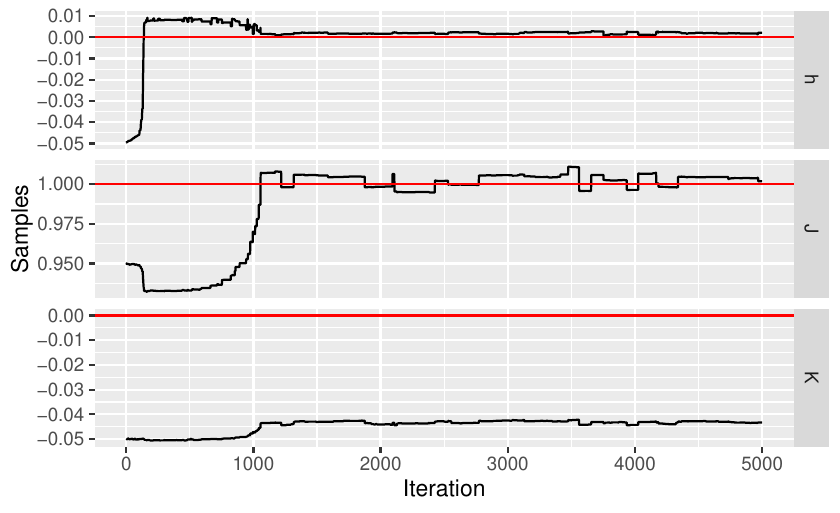}
                }
            \end{minipage}
            \begin{minipage}{0.32\textwidth}
                \centering
            \subfigure[RMAHMC]{\label{uniRMAHMC_crit}
                    \includegraphics[width=\textwidth]{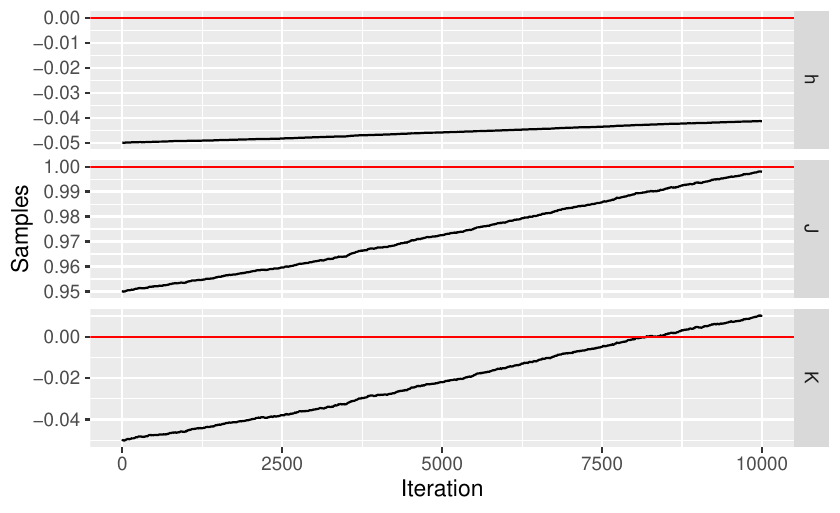}
                }
            \end{minipage}
            \begin{minipage}{0.32\textwidth}
                \centering
        \subfigure[RMAHMC-AMH]{\label{uniHybrid_crit}
                    \includegraphics[width=\textwidth]{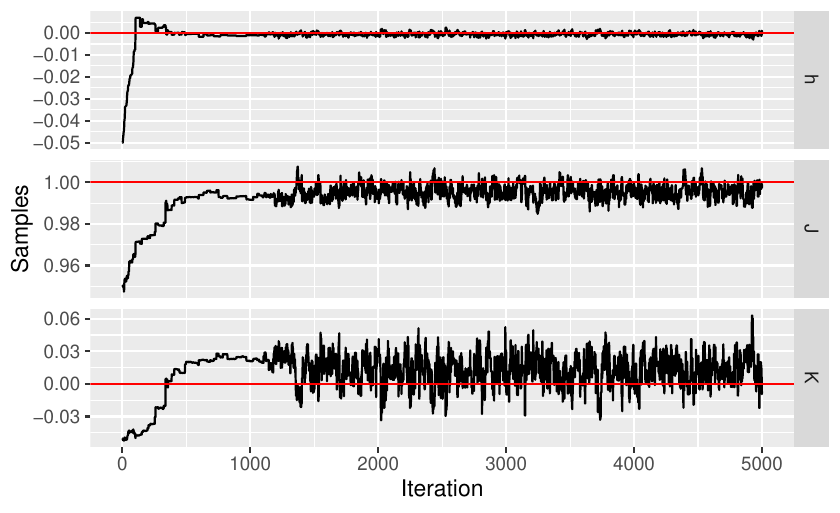}
                }
            \end{minipage}
        \end{minipage}
    }
    \caption{Trace plots associated with MCMC samples from the posterior distribution of $(K,J,h)$ based on $N = 300$ and $M=1000$ sampled configurations. The first row displays the three MCMC algorithms for $(K,J,h)=(0.5,0.3,0.1)$ and the second row for $(K,J,h)=(0,1,0)$.}
    \label{TraceK=0.5J=1}
\end{figure}

Trace plots in Figure \ref{TraceK=0.5J=1} show that for $(K,J,h)=(0.5,0.3,0.1)$, both the Adaptive Metropolis--Hastings (AMH) and Riemannian Manifold HMC (RMAHMC) samplers achieve reasonable mixing, but the hybrid RMAHMC--AMH algorithm attains faster convergence and reduced autocorrelation.  At the critical point $(K,J,h)=(0,1,0)$, however, difficulties are more pronounced. The AMH chain often becomes trapped far from the true $K$ value (see Figures \ref{uniAHMC_crit}), while RMAHMC converges but only after very long runs (see Figures \ref{uniRMAHMC_crit}), consistent with the slow dynamics observed in related high-dimensional models \citep{Jacob_O’Leary_Atchadé_2020}. In contrast, the hybrid RMAHMC--AMH method again exhibits stable convergence and effective mixing (Figures \ref{uniAHMC_crit}--\ref{uniHybrid_crit}).  

%%%%%%%%%%%%%%%%%%%%%%%%%%%%%

\begin{figure}[H]
    \makebox[\textwidth][c]{ % Centering and allowing overflow
        \subfigure[$(0.5,0.3,0.1)$]{\label{denuni}
            \includegraphics[width=7.9cm]{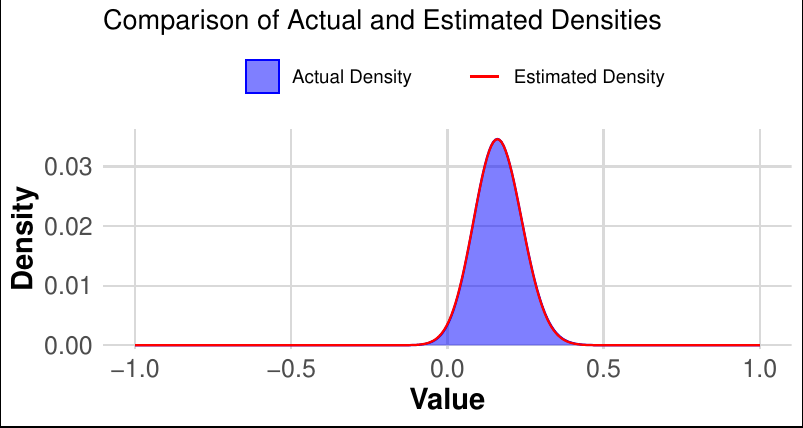}
        }
        \subfigure[$(0,1,0)$]{\label{dencritical}
            \includegraphics[width=7.9cm]{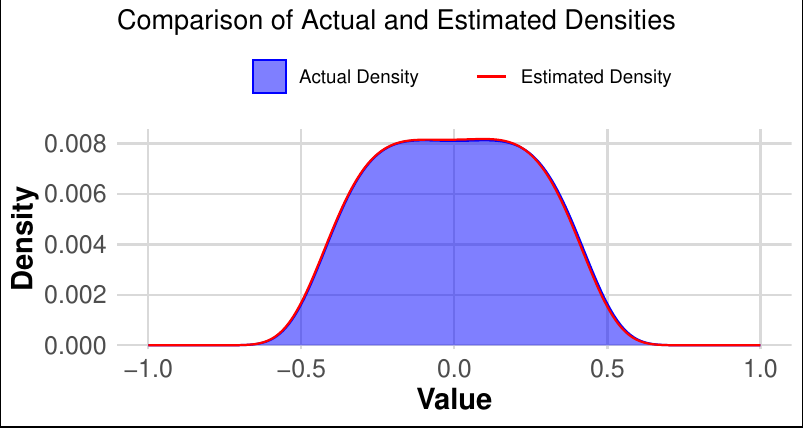}
        }   
    }
\caption{Boltzmann-Gibbs distribution of $m_N(\mathbf{x})$ at $N = 300$, with $M = 1000$ for the true and recovered parameters at (a) $\boldsymbol{\theta}=(0.5,0.3,0.1)$ and (b) $\boldsymbol{\theta}=(0,1,0)$. In (a), the peak of the distribution is centered around  $Em_N(\mathbf{x})$ for fixed $N$ and  corresponds to a global stable state of the system while (b) has a flat mode, around  $Em_N(\mathbf{x})=0$ as $N$ increases.%, indicating a certain level of uncertainty or lack of specificity in the system, as different values are equally probable within the flat region..
}\label{den_uni}
\end{figure}

Finally, Figure \ref{den_uni} compares density estimates of $m_N(\mathbf{x})$ obtained from posterior means of $\bm{\theta}$ using hybrid RMAHMC--AMH samples. Even at the critical point $(K,J,h)=(0,1,0)$, the estimated densities closely match the true distribution (Figure \ref{dencritical}), underscoring the robustness of the hybrid approach.

\subsection{Case 3: Non-identifiable Likelihood Surface} \label{sec:nonident}
The final scenario concerns parameter regimes where the inversion problem is \emph{non-identifiable}. In this setting, multiple combinations of $(K,J,h)$ yield nearly indistinguishable likelihoods, producing an effectively flat likelihood surface in certain directions of parameter space. Although we were unable to characterize the subspace of $\bm{\theta}$ leading to this degeneracy, such behavior is most pronounced when all three parameters take relatively large values. Importantly, non-identifiability is not restricted to multimodal cases: it can also occur in unimodal regimes. As an illustration, we examine $\bm{\theta}=(0.5,0.3,0.9)$, for which the magnetization distribution remains unimodal.

\begin{figure}[H]
    \centering
    \makebox[\textwidth][c]{ % Allow images to extend beyond margins
        \begin{minipage}{1.2\textwidth}
            \centering
            % First row (2 images)
            \begin{minipage}{0.48\textwidth}
                \centering
                \subfigure[]{\label{heatmapNon}
                    \includegraphics[width=\textwidth]{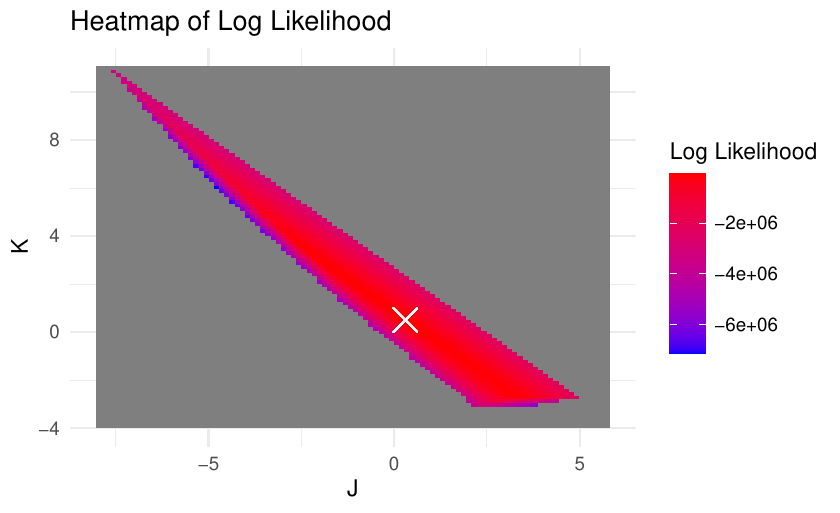}
                }
            \end{minipage}
            \begin{minipage}{0.48\textwidth}
                \centering
                \subfigure[]{\label{TraceNonident}
                    \includegraphics[width=\textwidth]{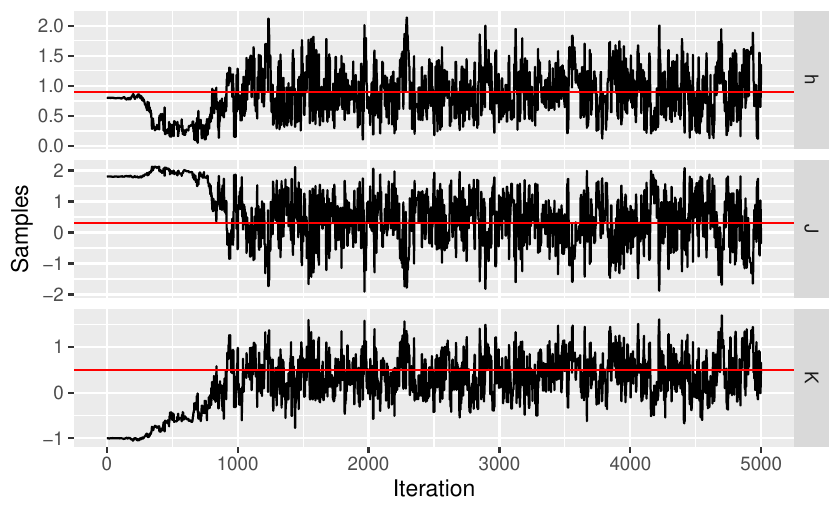}
                }
            \end{minipage} \\[1em] % Space between rows

            % Second row (2 images)
            \begin{minipage}{0.48\textwidth}
                \centering
                \subfigure[]{\label{GibbsB}
                    \includegraphics[width=\textwidth]{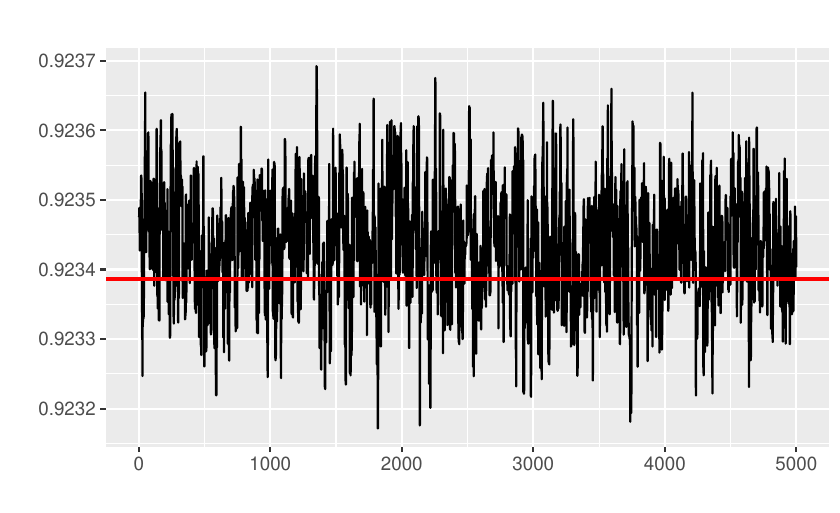}
                }
            \end{minipage}
            \begin{minipage}{0.48\textwidth}
                \centering
                \subfigure[]{\label{densityNonident}
                    \includegraphics[width=\textwidth]{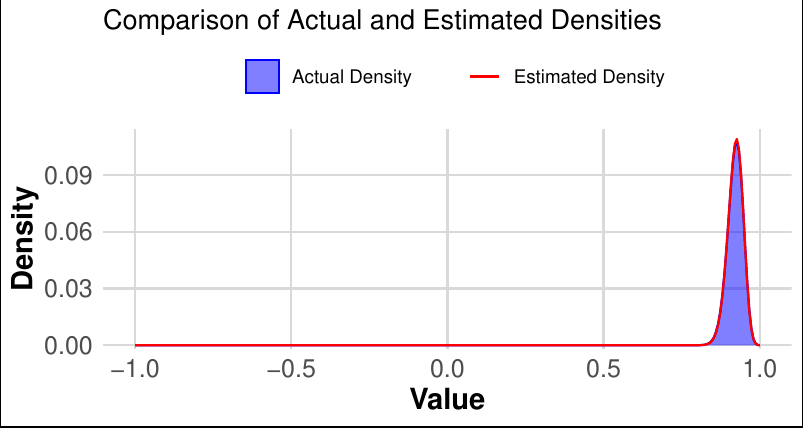}
                }
            \end{minipage}
        \end{minipage}
    }
    \caption{$K=0.5, J=0.3$ and $h=0.9$. Plot (a) is the heatmap of the log-likelihood as a function of $K$ and $J$ for fixed $h=0.9$. Plot (b) displays the trace plot of the hybrid RMHMC-AMH for parameter sampling from the posterior distribution at $N = 300$ and $M=1,000$ sampled configurations. Plot (c) depicts the theoretical mean as a function of the recovered parameters and (d) is the associated density plot.}
    \label{GibbsNonident}
\end{figure}

Figure \ref{TraceNonident} shows MCMC traces obtained with the hybrid RMAHMC--AMH sampler (trace plots for AMH and RMAHMC alone are given in Appendix \ref{AMH RMAHMC}). The chain converges toward the true parameters within approximately $2{,}000$ iterations, but the resulting credible intervals are substantially wider than in the bimodal or unimodal cases, reflecting weak identifiability. This effect is further illustrated in Figure \ref{GibbsB}, which tracks the theoretical mean magnetization $\mu=\mu_1(K,J,h)$. Unlike the raw parameters, $\mu$ is recovered with less variability and still captures the truth, underscoring that identifiability issues manifest more strongly at the parameter level than at the level of derived macroscopic quantities.

Finally, Figure \ref{densityNonident} demonstrates that the estimated density of $m_N(\mathbf{x})$, computed from posterior means of $\bm{\theta}$, remains in close agreement with the true density. Thus, while individual parameters exhibit substantial posterior uncertainty, the induced distribution of observables is nonetheless well captured—a hallmark of non-identifiable models.
%%%%%%%%%%%%%%%%%%%%%%%%%%%%%

%%%%%%%%%%%%%%%%%%%%%%%%%%%
\subsection{Convergence Measures}\label{convergence.measures}
While the trace plots are visually informative, it is useful to report a quantitative convergence diagnostic. We use the Gelman--Rubin statistic $\hat{R}$ \citep{gelman1992inference}. The trace plots for all five scenarios were shown earlier using 5,000 samples to illustrate qualitative trends. Here we run the sampler for 100,000 iterations and compute $\hat{R}$ based on sample sizes of 5,000, 50,000, and 100,000, with burn-in lengths of 2,500, 40,000, and 90,000 respectively. Table~\ref{tab:gelman_rubin_k} reports results for parameter $K$ only, since those for $J$ and $h$ are very similar.  
\begin{table}[h]
\centering
\caption{Gelman--Rubin statistic $\hat{R}$ for parameter $K$. Values close to 1 indicate convergence.}
\label{tab:gelman_rubin_k}
\begin{tabularx}{\textwidth}{XXXXX}
\toprule
\multirow{2}{*}{Scenario ($K, J, h)$} & \multirow{2}{*}{Sampler} & \multicolumn{3}{c}{Length of Run} \\
\cmidrule{3-5}
 &  & 5{,}000 & 50{,}000 & 100{,}000 \\
\midrule
$(0,1.2,0)$ & AMH    & 2.26  & 1.33   & 1.03 \\
            & RMAHMC & 10.69 & 1.124  & 1.034 \\
            & Hybrid & 1.009 & 1.0009 & 1.0009 \\
\addlinespace
$(1.67,0.01,0.1)$ & AMH    & 1.18  & 1.04   & 1.001 \\
                  & RMAHMC & 2.02  & 1.007  & 1.05 \\
                  & Hybrid & 1.01  & 1.001  & 1.001 \\
\addlinespace
$(0.5,0.3,0.1)$ & AMH    & 1.08  & 1.01   & 1.01 \\
                & RMAHMC & 1.66  & 1.01   & 1.01 \\
                & Hybrid & 1.008 & 1.002  & 1.001 \\
\addlinespace
$(0,1,0)$ & AMH    & 8.44  & 7.55   & 1.96 \\
          & RMAHMC & 20.98 & 1.07   & 1.006 \\
          & Hybrid & 1.01  & 1.002  & 1.0007 \\
\addlinespace
$(0.5,0.3,0.9)$ & AMH    & 4.75  & 1.79   & 2.63 \\
                & RMAHMC & 1.005 & 1.007  & 1.003 \\
                & Hybrid & 1.0001 & 1.0008 & 1.003 \\
\bottomrule
\end{tabularx}
\end{table}
Values of $\hat{R}$ close to 1 indicate convergence of the Markov chain to the stationary distribution, while values above about 1.1 typically signal lack of convergence \citep{gelman1992inference}. Table~\ref{tab:gelman_rubin_k} shows that all three samplers eventually converge for sufficiently long runs. However, even with 100,000 samples the AMH sampler exhibits poor convergence in the $(0,1,0)$ and $(0.5,0.3,0.9)$ scenarios. By contrast, the Hybrid sampler converges rapidly and consistently across all cases.  

\subsection{Simulation Study}
To evaluate the RMHMC--AMH algorithm’s ability to recover parameters of the true data-generating process, we conducted a small simulation study. One hundred datasets were generated from each of the five $(K,J,h)$ configurations described in Section~\ref{test}. For each dataset, 95\% credible intervals for $(K,J,h)$ were estimated using 5,000 posterior samples from the RMHMC--AMH algorithm, discarding the first 2,500 as burn-in. This sample size is sufficient given the rapid convergence and good mixing properties of RMHMC--AMH.  

We report empirical coverage probabilities (the proportion of intervals containing the true parameter value) and average credible interval widths. Results are shown in Table~\ref{tab:alpha_models}. In all identifiable cases, coverage is close to the nominal 95\%. The exception is parameter $J$ near the phase transition boundary, where coverage drops slightly to 90\% due to narrower intervals. In the non-identifiable case, coverage reaches 100\% but only because the intervals are extremely wide, reflecting structural non-identifiability. The only other case whose coverage was further from 0.95 than expected was the parameter $J$ corresponding to the  point where phase transition begins. Rather than 0.95, the coverage was 0.9.  Notice that for this case the intervals were much narrower than for the unimodal 1 case.  Overall, the study confirms that RMHMC--AMH produces reliable inference across challenging parameter regimes of the Ising model with a three-body interaction.  

\begin{table}[H]
 \centering
 \caption{Simulation study results. Coverage is the empirical probability that the 95\% credible interval contains the true value; width is the average interval width.}
\begin{tabular}{l l l l } \toprule
Case & \multicolumn{1}{c}{$\bm{\theta} = (K, J, h)$} & \multicolumn{1}{c}{Coverage} & \multicolumn{1}{c}{Width} \\
\midrule 
Bimodal 1        & (1.67, 0.01, 0.10) & (0.95, 0.96, 0.97) & (0.224, 0.284, 0.077) \\
Bimodal 2        & (0.00, 1.20, 0.00) & (0.98, 0.94, 0.98) & (0.042, 0.019, 0.006) \\
Unimodal 1       & (0.50, 0.30, 0.10) & (0.97, 0.97, 0.99) & (0.990, 0.373, 0.033) \\
Unimodal 2       & (0.00, 1.00, 0.00) & (0.96, 0.90, 0.95) & (0.055, 0.014, 0.003) \\
non-identifiable  & (0.50, 0.30, 0.90) & (1.00, 1.00, 1.00) & (1.610, 2.813, 1.367) \\
\bottomrule
\end{tabular}
 \label{tab:alpha_models}
 \end{table}

\section{Conclusion}\label{sec:conclusion}
We studied the Bayesian inverse problem for the mean-field Ising model augmented with three-body interactions.  Concretely, given observations \(\mathbf{X}\) (a collection of i.i.d.\ configurations) we targeted the posterior $p(\bm\theta | \mathbf{X})$ and developed both analytical and computational tools to perform inference when the likelihood is irregular due to higher-order interactions.

Our methodological contributions are threefold and are summarised here succinctly.  First, using statistical-mechanics expansions of the partition function we provided explicit expressions used to compute the log-likelihood, its gradient and (when feasible) the Fisher information.  Second, to overcome severe parameter correlations and poor mixing we designed a hybrid sampler that alternates geometry-aware Riemann--manifold HMC steps with adaptive Metropolis proposals (RMAHMC--AMH).  The alternation preserves $p(\bm\theta | \mathbf{X})$ and combines principled local moves (using the metric $G(\bm\theta)$) with occasional larger adaptive jumps that help traverse flat or multimodal regions.  Third, we carefully addressed numerical issues that arise in practice: grid-based initialisation to place chains in high-density regions, shrinkage/regularisation of $G(\bm\theta)$ to avoid numerical singularity, and an annealing schedule for the regulariser to recover the true geometry after burn-in.

Numerically, across representative regimes (bimodal, unimodal, and near-non-identifiable) the hybrid RMAHMC--AMH sampler outperformed stand-alone AMH and RMAHMC in terms of rapid convergence and effective sample quality.  Convergence diagnostics (Gelman--Rubin) and a small simulation study (empirical coverage of 95\% credible intervals) corroborate these observations: identifiable cases exhibit near-nominal coverage and moderate interval widths, whereas structurally non-identifiable regimes produce wide intervals that nevertheless correctly capture the induced observable distribution.  Analytically, non-identifiability was signalled by near-singularity of \(I_M(\bm\theta)\); equivalently, when \(\lambda_{\min}(I_M(\bm\theta))\) is small, the quadratic approximation of the log-likelihood is nearly flat along the corresponding eigendirections and parameter recovery becomes inherently ill-posed.

The results have several implications for machine learning, in particular for energy-based and Boltzmann--machine learning.  First, the techniques developed here apply directly to learning higher-order Boltzmann machines: the posterior sampling framework provides principled uncertainty quantification for weights (including three-body terms), an advantage over point-estimate methods such as contrastive divergence.  Second, the Riemannian preconditioning we used (metric $G(\bm\theta)$) suggests practical improvements for gradient-based training of energy-based models: local curvature information can stabilise and accelerate learning in parameter directions with strong correlation.  Third, the identifiability diagnostics (small \(\lambda_{\min}\), wide credible intervals, but accurate predictive densities) warn practitioners that good predictive performance does not necessarily imply that individual parameters are well identified — an important consideration when learned interactions are interpreted or used downstream.

There are natural and concrete directions for future work.  From a computational perspective, scaling to larger systems and higher-dimensional parameter spaces will require stochastic/mini-batch versions of RMAHMC or low-rank approximations of \(G(\bm\theta)\), as well as more efficient estimators of high-order moments.  From a modelling viewpoint, extending the theoretical analysis and inference to non-mean-field graphs, to sparse or structured higher-order interactions, and to partially observed systems (missing nodes or hidden variables) is both important and challenging.  Finally, methodological extensions that combine our posterior-sampling approach with variational approximations or tempering schemes (multiple temperatures / multi–\(\beta\) observations) may mitigate identifiability issues by increasing the information content of the data.

In summary, this work combines statistical-mechanics expansions, a carefully regularised Riemannian sampling strategy, and practical computational prescriptions to enable robust Bayesian inference for three-body interacting Ising models.  The methods are immediately relevant to research in energy-based models and Boltzmann-machine learning, where higher-order interactions and principled uncertainty quantification are increasingly of interest.

\begin{appendices}

\section{AMH and RMAHMC: non-identifiable case}\label{AMH RMAHMC}

\begin{figure}[H]
\makebox[\textwidth][c]{ % Centering and allowing overflow
\subfigure[AMH]{\label{noniden_amh}
\includegraphics[width=7.9cm]{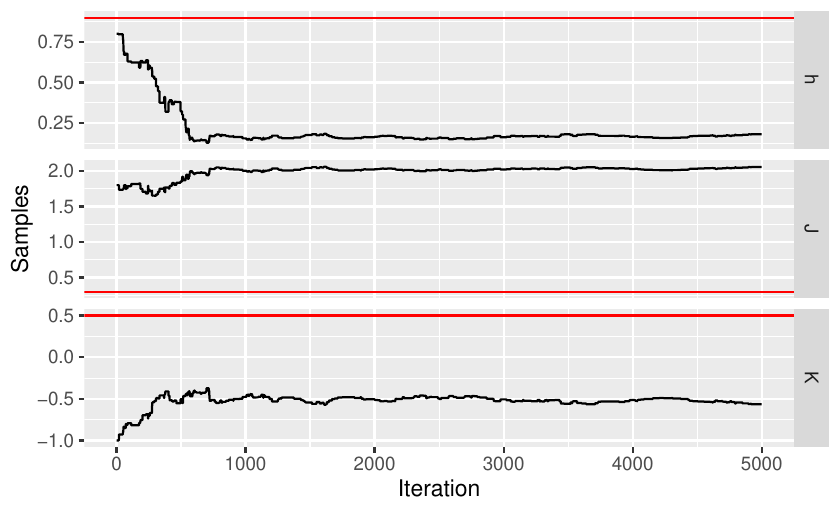}
}
\subfigure[RMAHMC]{\label{noniden_rmahmc}
\includegraphics[width=7.9cm]{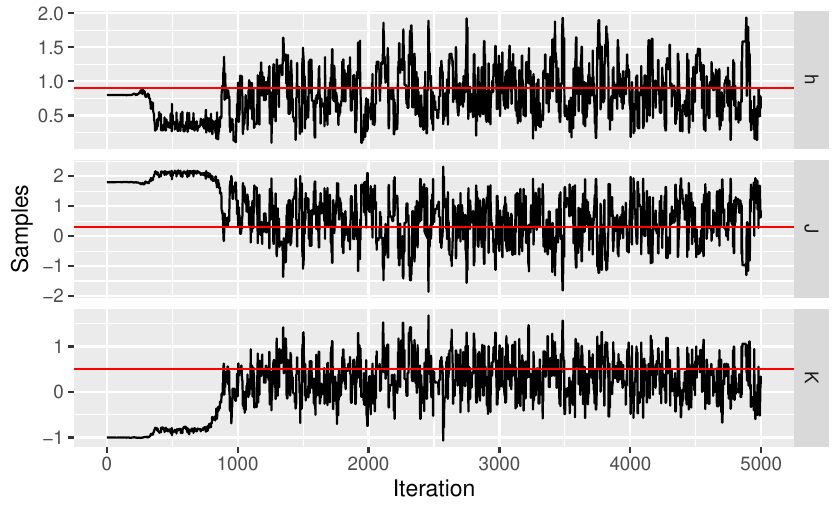}
}
}
\caption{Trace plot associated to the AMH and RMAHMC algorithm for the recovery of $(K,J,h)=(0.5,0.3,0.9)$ with $N=300$ and $M=1000$ sampled configurations.
}\label{noniden_AR}
\end{figure}

\end{appendices}

\singlespacing
\bibliographystyle{jasa3}
%\bibliography{refs}

\begin{thebibliography}{41}
\newcommand{\enquote}[1]{``#1''}
\expandafter\ifx\csname natexlab\endcsname\relax\def\natexlab#1{#1}\fi
\expandafter\ifx\csname url\endcsname\relax
  \def\url#1{{\tt #1}}\fi
\expandafter\ifx\csname urlprefix\endcsname\relax\def\urlprefix{URL }\fi

\bibitem[\protect\citeauthoryear{Ackley, Hinton, and Sejnowski}{Ackley
  et~al.}{1985}]{Ackley1985}
Ackley, D.~H., Hinton, G.~E., and Sejnowski, T.~J. (1985), \enquote{A learning
  algorithm for Boltzmann machines,} {\em Cognitive Science\/}, 9, 147--169.

\bibitem[\protect\citeauthoryear{Alvarez-Rodriguez, Battiston, de~Arruda,
  Moreno, Perc, and Latora}{Alvarez-Rodriguez
  et~al.}{2021}]{AlvarezBattiston2021}
Alvarez-Rodriguez, U., Battiston, F., de~Arruda, G.~F., Moreno, Y., Perc, M.,
  and Latora, V. (2021), \enquote{Evolutionary dynamics of higher-order
  interactions in social networks,} {\em Nature human behaviour\/}, 5,
  586–595, \urlprefix\url{http://dx.doi.org/10.1038/s41562-020-01024-1}.

\bibitem[\protect\citeauthoryear{Aurell and Ekeberg}{Aurell and
  Ekeberg}{2012}]{Aurell_2012}
Aurell, E. and Ekeberg, M. (2012), \enquote{Inverse {I}sing Inference Using All
  the Data,} {\em Phys. Rev. Lett.\/}, 108, 090201,
  \urlprefix\url{https://link.aps.org/doi/10.1103/PhysRevLett.108.090201}.

\bibitem[\protect\citeauthoryear{Battiston, Amico, Barrat, Bianconi, Ferraz~de
  Arruda, Franceschiello, Iacopini, Kéfi, Latora, Moreno, Murray, Peixoto,
  Vaccarino, and Petri}{Battiston et~al.}{2021}]{BattistonAmico2021}
Battiston, F., Amico, E., Barrat, A., Bianconi, G., Ferraz~de Arruda, G.,
  Franceschiello, B., Iacopini, I., Kéfi, S., Latora, V., Moreno, Y., Murray,
  M.~M., Peixoto, T.~P., Vaccarino, F., and Petri, G. (2021), \enquote{The
  physics of higher-order interactions in complex systems,} {\em Nature
  physics\/}, 17, 1093–1098,
  \urlprefix\url{http://dx.doi.org/10.1038/s41567-021-01371-4}.

\bibitem[\protect\citeauthoryear{Baxter}{Baxter}{2016}]{baxter2016exactly}
Baxter, R.~J. (2016), {\em Exactly Solved Models in Statistical Mechanics\/},
  Elsevier,
  \urlprefix\url{https://www.elsevier.com/books/exactly-solved-models-in-statistical-mechanics/baxter/978-0-12-083180-7}.
  Reprint of the 1982 original by Academic Press.

\bibitem[\protect\citeauthoryear{Benson, Abebe, Schaub, Jadbabaie, and
  Kleinberg}{Benson et~al.}{2018}]{BensonAbebe2018}
Benson, A.~R., Abebe, R., Schaub, M.~T., Jadbabaie, A., and Kleinberg, J.
  (2018), \enquote{Simplicial closure and higherorder link prediction,} {\em
  Proc. Natl Acad. Sci. USA\/}, 115, E11221–E11230,
  \urlprefix\url{http://dx.doi.org/10.1073/pnas.1800683115}.

\bibitem[\protect\citeauthoryear{Bhattacharya and Mukherjee}{Bhattacharya and
  Mukherjee}{2018}]{Bhattacharya_MPLE_2018}
Bhattacharya, B.~B. and Mukherjee, S. (2018), \enquote{Inference in {I}sing
  models,} {\em Bernoulli: official journal of the Bernoulli Society for
  Mathematical Statistics and Probability\/}, 24, 493–525,
  \urlprefix\url{http://dx.doi.org/10.3150/16-bej886}.

\bibitem[\protect\citeauthoryear{Carbone}{Carbone}{2025}]{Carbone2025}
Carbone, D. (2025), \enquote{Hitchhiker's guide on the relation of Energy-Based
  Models with other generative models, sampling and statistical physics: a
  comprehensive review,} {\em Transactions on Machine Learning Research\/}.

\bibitem[\protect\citeauthoryear{Chatterjee}{Chatterjee}{2007}]{Chatterjee_2007}
Chatterjee, S. (2007), \enquote{Estimation in spin glasses: A first step,} {\em
  Annals of statistics\/}, 35, 1931–1946,
  \urlprefix\url{http://dx.doi.org/10.1214/009053607000000109}.

\bibitem[\protect\citeauthoryear{Contucci, Kertész, and Osabutey}{Contucci
  et~al.}{2022}]{Contucci_Kertész_Osabutey_2022}
Contucci, P., Kertész, J., and Osabutey, G. (2022), \enquote{Human-{AI}
  ecosystem with abrupt changes as a function of the composition,} {\em PloS
  one\/}, 17, e0267310,
  \urlprefix\url{http://dx.doi.org/10.1371/journal.pone.0267310}.

\bibitem[\protect\citeauthoryear{Contucci, Mingione, and Osabutey}{Contucci
  et~al.}{2024}]{Contucci_Mingione_Osabutey_2023}
Contucci, P., Mingione, E., and Osabutey, G. (2024), \enquote{Limit theorems
  for the cubic mean-field Ising model,} in {\em Annales Henri Poincar{\'e}\/},
  Springer.

\bibitem[\protect\citeauthoryear{Contucci, Osabutey, and Vernia}{Contucci
  et~al.}{2023}]{Contucci_Osa_Ver_2023}
Contucci, P., Osabutey, G., and Vernia, C. (2023), \enquote{Inverse problem
  beyond two-body interaction: The cubic mean-field {I}sing model,} {\em Phys.
  Rev. E\/}, 107, 054124,
  \urlprefix\url{https://link.aps.org/doi/10.1103/PhysRevE.107.054124}.

\bibitem[\protect\citeauthoryear{Decelle, de~Jes\'us Navas~G\'omez, and
  Seoane}{Decelle et~al.}{2025}]{DecelleNavasGomezSeoane2025}
Decelle, A., de~Jes\'us Navas~G\'omez, A., and Seoane, B. (2025),
  \enquote{Inferring High-Order Couplings with Neural Networks,} {\em arXiv
  preprint arXiv:2501.06108\/}.

\bibitem[\protect\citeauthoryear{Decelle and Ricci-Tersenghi}{Decelle and
  Ricci-Tersenghi}{2016}]{DecelleR2016}
Decelle, A. and Ricci-Tersenghi, F. (2016), \enquote{Solving the inverse Ising
  problem by mean-field methods in a clustered phase space with many states,}
  {\em Physical review. E\/}, 94, 012112,
  \urlprefix\url{http://dx.doi.org/10.1103/PhysRevE.94.012112}.

\bibitem[\protect\citeauthoryear{Dembo and Zeitouni}{Dembo and
  Zeitouni}{2010}]{Dembo_Zeitouni_2010}
Dembo, A. and Zeitouni, O. (2010), {\em Large Deviations Techniques and
  Applications\/}, Berlin, Germany: Springer.

\bibitem[\protect\citeauthoryear{Ellis}{Ellis}{2006}]{Ellis06}
Ellis, R.~S. (2006), {\em Entropy, large deviations, and statistical
  mechanics\/}, Classics in Mathematics, Springer-Verlag, Berlin,
  \urlprefix\url{https://doi.org/10.1007/3-540-29060-5}. Reprint of the 1985
  original.

\bibitem[\protect\citeauthoryear{Fedele and Vernia}{Fedele and
  Vernia}{2017}]{Fedele_Vernia_2017}
Fedele, M. and Vernia, C. (2017), \enquote{Inverse problem for multispecies
  ferromagneticlike mean-field models in phase space with many states,} {\em
  Physical review. E\/}, 96, 042135,
  \urlprefix\url{http://dx.doi.org/10.1103/PhysRevE.96.042135}.

\bibitem[\protect\citeauthoryear{Fedele, Vernia, and Contucci}{Fedele
  et~al.}{2013}]{FedeleVernia2013}
Fedele, M., Vernia, C., and Contucci, P. (2013), \enquote{Inverse problem
  robustness for multi-species mean-field spin models,} {\em J. of Phys. A:
  Math. and Theo.\/}, 46, 065001,
  \urlprefix\url{https://doi.org/10.1088/1751-8113/46/6/065001}.

\bibitem[\protect\citeauthoryear{Gallo, Barra, and Contucci}{Gallo
  et~al.}{2009}]{GalloBarra2009}
Gallo, I., Barra, A., and Contucci, P. (2009), \enquote{Parameter evaluation of
  a simple mean-field model of social interaction,} {\em Mathematical Models
  and Methods in Applied Sciences\/}, 19, 1427--1439.

\bibitem[\protect\citeauthoryear{Gelman and Rubin}{Gelman and
  Rubin}{1992}]{gelman1992inference}
Gelman, A. and Rubin, D.~B. (1992), \enquote{Inference from iterative
  simulation using multiple sequences,} {\em Statistical science\/}, 7,
  457--472.

\bibitem[\protect\citeauthoryear{Girolami and Calderhead}{Girolami and
  Calderhead}{2011}]{Girolami_Calderhead_2011}
Girolami, M. and Calderhead, B. (2011), \enquote{Riemann manifold Langevin and
  Hamiltonian Monte Carlo methods: Riemann Manifold Langevin and Hamiltonian
  Monte Carlo Methods,} {\em Journal of the Royal Statistical Society. Series
  B, Statistical methodology\/}, 73, 123–214,
  \urlprefix\url{http://dx.doi.org/10.1111/j.1467-9868.2010.00765.x}.

\bibitem[\protect\citeauthoryear{Goldstone}{Goldstone}{2015}]{political_rev2015}
Goldstone, J.~A. (2015), {\em The encyclopedia of political revolutions\/},
  Routledge.

\bibitem[\protect\citeauthoryear{Gu and Zhang}{Gu and
  Zhang}{2022}]{GuZhang2022}
Gu, J. and Zhang, K. (2022), \enquote{Thermodynamics of the Ising model encoded
  in restricted Boltzmann machines,} {\em Entropy\/}, 24, 1701.

\bibitem[\protect\citeauthoryear{Habeck}{Habeck}{2014}]{Habeck_Bayes2014}
Habeck, M. (2014), \enquote{Bayesian approach to inverse statistical
  mechanics,} {\em Phys. Rev. E\/}, 89, 052113,
  \urlprefix\url{https://link.aps.org/doi/10.1103/PhysRevE.89.052113}.

\bibitem[\protect\citeauthoryear{Hamilton, Leskovec, and Jurafsky}{Hamilton
  et~al.}{2016}]{hamilton2016cultural}
Hamilton, W.~L., Leskovec, J., and Jurafsky, D. (2016), \enquote{Cultural shift
  or linguistic drift? comparing two computational measures of semantic
  change,} in {\em Proceedings of the conference on empirical methods in
  natural language processing. Conference on empirical methods in natural
  language processing\/}, volume 2016, NIH Public Access.

\bibitem[\protect\citeauthoryear{Hinton and Sejnowski}{Hinton and
  Sejnowski}{1986}]{HintonSejnowski1986}
Hinton, G.~E. and Sejnowski, T.~J. (1986), \enquote{Learning and relearning in
  Boltzmann machines,} in Rumelhart, D.~E. and McClelland, J.~L. (editors),
  {\em Parallel Distributed Processing: Explorations in the Microstructure of
  Cognition, Vol. 1: Foundations\/}, MIT Press, 282--317.

\bibitem[\protect\citeauthoryear{Inglehart}{Inglehart}{2020}]{inglehart2020modernization}
Inglehart, R. (2020), {\em Modernization and postmodernization: Cultural,
  economic, and political change in 43 societies\/}, Princeton university
  press.

\bibitem[\protect\citeauthoryear{Ito and Kohring}{Ito and
  Kohring}{1994}]{Ito_Kohring_1994}
Ito, N. and Kohring, G.~A. (1994), \enquote{Single-spin algorithms- Which are
  more efficient?} {\em Intl. J. Mod. Phys. C\/}, 05.

\bibitem[\protect\citeauthoryear{Jacob, O’Leary, and Atchadé}{Jacob
  et~al.}{2020}]{Jacob_O’Leary_Atchadé_2020}
Jacob, P.~E., O’Leary, J., and Atchadé, Y.~F. (2020), \enquote{Unbiased
  Markov chain Monte Carlo methods with couplings,} {\em Journal of the Royal
  Statistical Society. Series B, Statistical methodology\/}, 82, 543–600,
  \urlprefix\url{http://dx.doi.org/10.1111/rssb.12336}.

\bibitem[\protect\citeauthoryear{Kim, Bhattacharya, and Maiti}{Kim
  et~al.}{2021}]{kim2021variational}
Kim, M., Bhattacharya, S., and Maiti, T. (2021), \enquote{Variational Bayes
  algorithm and posterior consistency of {I}sing model parameter estimation,} .

\bibitem[\protect\citeauthoryear{Mézard and Mora}{Mézard and
  Mora}{2009}]{Mezard_Mora_2009}
Mézard, M. and Mora, T. (2009), \enquote{Constraint satisfaction problems and
  neural networks: A statistical physics perspective,} {\em Journal of
  physiology, Paris\/}, 103, 107–113,
  \urlprefix\url{http://dx.doi.org/10.1016/j.jphysparis.2009.05.013}.

\bibitem[\protect\citeauthoryear{Nguyen and Berg}{Nguyen and
  Berg}{2012}]{Nguyen_Berg_2012}
Nguyen, H.~C. and Berg, J. (2012), \enquote{Mean-field theory for the inverse
  Ising problem at low temperatures,} {\em Physical review letters\/}, 109,
  050602, \urlprefix\url{http://dx.doi.org/10.1103/PhysRevLett.109.050602}.

\bibitem[\protect\citeauthoryear{Nguyen, Zecchina, and Berg}{Nguyen
  et~al.}{2017}]{Nguyen_Zecchina_Berg_2017}
Nguyen, H.~C., Zecchina, R., and Berg, J. (2017), \enquote{Inverse statistical
  problems: from the inverse {I}sing problem to data science,} {\em Advances in
  physics\/}, 66, 197–261,
  \urlprefix\url{http://dx.doi.org/10.1080/00018732.2017.1341604}.

\bibitem[\protect\citeauthoryear{O'Leary, Wang, and Jacob}{O'Leary
  et~al.}{2020}]{O’Leary_Wang_Jacob_2020}
O'Leary, J., Wang, G., and Jacob, P.~E. (2020), \enquote{Maximal Couplings of
  the {Metropolis-Hastings} Algorithm,} {\em arXiv preprint\/},
  \urlprefix\url{https://arxiv.org/abs/2010.08573}.

\bibitem[\protect\citeauthoryear{Opoku, Osabutey, and Kwofie}{Opoku
  et~al.}{2019}]{Opoku_Osabutey_Kwofie_2019}
Opoku, A.~A., Osabutey, G., and Kwofie, C. (2019), \enquote{Parameter
  evaluation for a statistical mechanical model for binary choice with social
  interaction,} {\em Journal of probability and statistics\/}, 2019, 1–10,
  \urlprefix\url{http://dx.doi.org/10.1155/2019/3435626}.

\bibitem[\protect\citeauthoryear{Rodriguez and Laio}{Rodriguez and
  Laio}{2014}]{Rodriguez_Laio_2014}
Rodriguez, A. and Laio, A. (2014), \enquote{Machine learning. Clustering by
  fast search and find of density peaks,} {\em Science (New York, N.Y.)\/},
  344, 1492–1496, \urlprefix\url{http://dx.doi.org/10.1126/science.1242072}.

\bibitem[\protect\citeauthoryear{Schneidman, Berry, Segev, and
  Bialek}{Schneidman et~al.}{2006}]{Schneidman2006}
Schneidman, E., Berry, M., Segev, R., and Bialek, W. (2006), \enquote{Weak
  pairwise correlations imply strongly correlated network states in a neural
  population,} {\em Nature\/}, 440, 1007--1012.

\bibitem[\protect\citeauthoryear{Skocpol}{Skocpol}{1979}]{skocpol1979states}
Skocpol, T. (1979), {\em States and social revolutions: A comparative analysis
  of France, Russia and China\/}, Cambridge University Press.

\bibitem[\protect\citeauthoryear{Subramanian and Lebowitz}{Subramanian and
  Lebowitz}{1999}]{SubLeo1999}
Subramanian, B. and Lebowitz, J. (1999), \enquote{The study of a three-body
  interaction Hamiltonian on a lattice,} {\em J. Phys. A: Math. Gen.\/}, 32.

\bibitem[\protect\citeauthoryear{Talagrand}{Talagrand}{2003}]{Talagrand_2003}
Talagrand, M. (2003), {\em Spin Glasses: A Challenge for Mathematicians-Cavity
  and mean-field Models\/}, Berlin: Springer.

\bibitem[\protect\citeauthoryear{Torquato}{Torquato}{2009}]{Torquato_2009}
Torquato, S. (2009), \enquote{Inverse optimization techniques for targeted
  self-assembly,} {\em Soft matter\/}, 5, 1157,
  \urlprefix\url{http://dx.doi.org/10.1039/b814211b}.

\end{thebibliography}

\end{document}